\documentclass[12pt]{article}
 \usepackage{epsfig}
 \usepackage {graphicx}
 \usepackage[latin1]{inputenc}
 \usepackage{amsfonts}
 \usepackage{amsthm}
 \usepackage{amssymb, latexsym, amsmath}
 \newcommand{\inc}{{\it i}}

 \newcommand{\be}{\begin{equation}}
 \newcommand{\ee}{\end{equation}}
 \newcommand{\ba}{\begin{eqnarray}}
 \newcommand{\ea}{\end{eqnarray}}
 \newcommand{\epsilonbold}{\mbox{{\boldmath $\epsilon$}}}
 
 \newcommand{\robold}{\mbox{{\boldmath $\vec \rho$}}}
 \newcommand{\erbold}{\mbox{{\boldmath $\vec r$}}}

 \newcommand{\chibold}{\mbox{{\boldmath $\chi$}}}

  \newcommand{\Vbold}{\mbox{\boldmath $\vec {\boldmath{\,V}}$}}

  \newcommand{\eRbold}{\mbox{{\boldmath $\vec{R}$}}}

\begin{document}
 \title{
                 ${{~~~}^{^{^{
        Published~in\,:\,~Celestial~Mechanics~and~Dynamical~Astronomy
        \,,~Vol.\,104\,,~pp.\,257-289~\,(2009)
                  }}}}$\\
                  ~\\
 {\Large{\textbf{Tidal torques. A critical review of some techniques
 \\}
            }}}
 \author{
 {\Large{Michael Efroimsky}}\\
 {\small{US Naval Observatory, Washington DC 20392 USA}}\\
 {\small{e-mail: ~michael.efroimsky @ usno.navy.mil~}}\\
 ~\\
 {{and}}\\
 ~\\
  {\Large{James G. Williams}}\\
 {\small{Jet Propulsion Laboratory, California Institute of Technology, Pasadena CA 91109 USA}}\\
 {\small{e-mail: ~james.g.williams @ jpl.nasa.gov~}}
 }
 \date{}

 \maketitle
 \begin{abstract}

 We review some techniques employed in the studies of torques due to
 bodily tides, and explain why the MacDonald formula for the tidal torque is valid
 only in the zeroth order of the eccentricity divided by the quality factor, while
 its time-average is valid in the first order. As a result, the formula cannot be
 used for analysis in higher orders of $\,e/Q\,$. This necessitates some corrections in
 the current theory of tidal despinning and libration damping (though the qualitative
 conclusions of that theory may largely remain correct).

 We demonstrate that in the case when the inclinations are small and the phase
 lags of the tidal harmonics are proportional to the frequency, the Darwin-Kaula
 expansion is equivalent to a corrected version of the MacDonald method. The
 latter method rests on the assumption of existence of one total double bulge.
 The necessary correction to MacDonald's approach would be to assert (following
 Singer 1968) that the phase lag of this integral bulge is not constant, but is
 proportional to the instantaneous synodal frequency (which is twice the
 difference between the evolution rates of the true anomaly and the sidereal
 angle). This equivalence of two descriptions becomes violated by a nonlinear
 dependence of the phase lag upon the tidal frequency. It remains unclear
 whether it is violated at higher inclinations.

 Another goal of our paper is to compare two derivations of a popular formula for the tidal despinning
 rate, and to emphasise that both are strongly limited to the case of a vanishing
 inclination and a certain (sadly, unrealistic) law of frequency-dependence of
 the quality factor $\,Q\,$ -- the law that follows from the phase lag being
 proportional to frequency. One of the said derivations is based on the MacDonald
 torque, the other on the Darwin torque. Fortunately, the second approach is
 general enough to accommodate both a finite inclination and the actual rheology.

 We also address the rheological models with the $Q$ factor scaling as the
 tidal frequency to a positive fractional power, and disprove the popular
 belief that these models introduce discontinuities into the equations
 and thus are unrealistic at low frequencies. Although such models indeed
 make the conventional expressions for the torque diverge at vanishing
 frequencies, the emerging infinities reveal not the impossible nature of
 one or another rheology, but a subtle flaw in the underlying mathematical model
 of friction. Flawed is the common misassumption that damping merely
 provides phase lags to the terms of the Fourier series for the tidal potential.
 A careful hydrodynamical treatment by Sir George Darwin (1879), with viscosity
 explicitly included, had demonstrated that the magnitudes of the terms, too,
 get changed -- a fine detail later neglected as ``irrelevant". Reinstating of
 this detail tames the fake infinities and rehabilitates the ``impossible"
 scaling law (which happens to be the actual law the terrestrial
 planets obey at low frequencies).

 Finally, we explore the limitations of the popular formula interconnecting
the quality factor and the phase lag. It turns out that, for low values of Q,
the quality factor is no longer equal to the cotangent of the lag.
 \end{abstract}


 \section{Prologue}
 ${\left.~~~~~~\,~~~~~~~~~~~~~~~~~~~~~~~~~~~~~~~~~~~~~~~~~~~~~~~~~~
 \,\right.}^{\mbox{\small \it When~it~shall~be~found~that~much~is~omitted,}}$~~\\
 ${\left.~~~~~~~~~~~~~~~~~~~~~~~~~~~~~~~~~~~~~~~~~~~~~~~~~~~~
 ~~~~~\,\right.}^{\mbox{\small \it let~it~not~be~forgotten~that~much~likewise~is~performed}}$\\
 ${\left.~~~~~~~~~~~~~~~~~~~~~~~~~~~~~~~~~~~~~~~~~~~~~~~~~~~~~~~~~~~~
 ~~~~~~~~~~~~~~~~~~~~~~~~~~~~~~\right.}^{
 \mbox{\small\it Samuel~Johnson, ~1755}
 }$\\
 ~\\
 In his short work {\emph{
 ``Untersuchung der Frage~...~,"}} known among the historians also as the
 {\emph{``Spin-Cycle essay,"}} Immanuel Kant (1754) stated that the Moon
 not only pulls the Earth, but also exerts a retarding torque upon its
 surface; this torque slows down the Earth's rotation and lets go only
 when terrestrial days become as long as lunar months. Although Kant
 had in mind only the ocean tides, not the bodily ones, we may say that,
 qualitatively, he predicted the celebrated $\,1:1\,$ spin-orbit resonance,
 the {\emph{pas de deux}} wherein Pluto and Charon are locked.

 For the first time, the idea of tidal action not being confined only to
 the fluid portion of the planet but affecting also the solid, so as to
 induce a state of varying strain, was put forward by John Herschel (son
 of astronomer William Herschel), as a minor aside in a paper devoted to
 volcanism and earthquakes (Herschel 1863). The earliest mathematical
 description of land tides in their dynamics was offered by George Darwin
 (son of naturalist Charles Darwin and great-grandson of poet and philosopher
 Erasmus Darwin).

 Following his predecessors Roche (1849) and Thompson (1863), who had
 calculated the figure of a static tide, Darwin (1879) assumed the Earth
 homogeneous and consisting of an incompressible fluid. To account for
 dynamics, he also assumed that the viscosity was the sole source of the
 tidal friction. Relying on this model, Darwin (1880, 1908) derived
 a tide-generated disturbing potential expanded into a Fourier series.
 Substitution thereof into the Lagrange-type planetary equations led
 him to expressions for the time derivatives of the orbital elements
 via partial derivatives of the disturbing potential with respect to
 the elements.

 An impressive generalisation of Darwin's work by Kaula (1964), and the
 subsequent flow of new concepts and applications (MacDonald 1964; Goldreich
 1966a,b; Goldreich \& Peale 1966; Singer 1968; Mignard 1979, 1980; Touma \&
 Wisdom 1994; Neron de Surgy \& Laskar 1997; Krasinsky 2002, 2006; Getino,
 Escapa \& García 2003; Ferraz Mello et al 2008; Efroimsky 2008) made bodily
 tides a rapidly developing area of the planetary
 astronomy. The vast and growing volume of the relevant material leaves one no
 chance to glean it all in one review. Therefore we shall concentrate on one
 special aspect of this research, the tidal torques emerging from the bodily
 tides. Moreover, we shall dwell solely on the techniques, not applications.

 Although our review will at times be very critical, it should from the beginning
 be agreed that our criticisms are intended in the spirit of the above quotation
 from Samuel Johnson.

 Along with reviewing the current state of the field, we shall provide some new
 results of our own. Specifically, we shall address the rheological models with
 the $Q$ factor scaling as the tidal frequency to a positive fractional
 exponential. We shall demonstrate that, contrary to the common opinion, such
 rheologies do {\emph{not}} cause infinities in the expression for the torque.
 We shall also derive an expression for the tidal torque decelerating a
 terrestrial planet obeying such a rheology. (That the realistic terrestrial
 bodies indeed obey this class of rheologies has been explained in Efroimsky \&
 Lainey 2007.)

 \section{Trivia}

 In this section, we shall briefly recall how a satellite-generated potential
 in a point on or inside the planet is expressed through the latitude,
 longitude, and the radial distance of the point.

 Let us begin  from the first principles. The dynamics of point masses
 $\,m_i\,$ located at inertial-frame-related positions $\,{\robold}_i\;$,
 \be
 m_i\;{
 \stackrel{\mbox{\bf{..}}}{\robold}
 }_i\;=\;m_i\; \sum_{j \neq
 i}\,\;G\,m_j\;\frac{{\bf {\erbold}}_{ij}}{r_{ij}^3}
 \;\;\;,\;\;\;\;\;{\bf {\erbold}}_{ij}\,\equiv\,{\bf
 {\robold}}_j\,-\,{\bf {\robold}}_i\;\;,
 \;\;\;\;i,j=1,...,N\;\;,\;\;\;
 \label{1}
 \ee
 may be conveniently reformulated in terms of the relative-to-the-primary locations
 \be
 \erbold_i\;\equiv\;\erbold_{0i}\;\equiv\;{\bf {\robold}}_i\;-\;{\bf {\robold}}_0\;\;\;,
 \label{2}
 \ee
 ${\bf {\robold}}_0\;$ standing for the position of the primary. The
 difference between
 \be
 {\stackrel{\mbox{\bf{..}}}{\robold}}_i\;=\;\sum_{j\neq i,0}\;\;G\;
 \frac{m_j\;{\erbold}_{ij}}{r_{ij}^3}\;+\;G\;\frac{m_0\;{\bf
 {\erbold}}_{i0}}{r_{i0}^3}\;
 \label{3}
  \ee
 and
 \be
  { \stackrel{\mbox{\bf{..}}}{\robold}}_0\;=\;\sum_{j\neq i, 0}\;\;G\;
 \frac{m_j\;{\bf {\erbold}}_{0j}}{r_{0j}^3}\;+\;G\;\frac{m_i\;{\bf
 {\erbold}}_{0i}}{r_{0i}^3}\;
 \label{4}
 \ee
 amounts to:
 \be
 {\mbox{\boldmath$\ddot {\erbold}$}}_i\;=\;
 \sum_{j\neq i,0}\;\;G\;\frac{m_j\;{\bf {\erbold}}_{ij}}{r_{ij}^3}
 \;-\;
 \sum_{j\neq i,0}\;\;G\;\frac{m_j\;{\erbold}_{j}}{r_{j}^3}
 \;-\;
 G\;\frac{\left(m_i\,+\,m_0\right)\;{\erbold}_{i}}{r_{i}^3}
 \;=\;-\;
 \frac{\partial\,U_{i}}{\partial\,{\erbold}_{i}}
 \label{5}
 \ee
 $U_i\;$ being the potential:
 \be
 {U}_{i}\;\equiv\;-\;\frac{G\;\left(m_i\;+\;m_0\right)}{r_i}\;+\;W_i\;\;,
 \label{6}
 \ee
with the disturbance
 \be
 W_i\;\equiv\;-\;\sum_{j\neq i}\;\;G\;m_j\;\left\{\frac{1}{r_{ij}}\;
 -\;\frac{{{\erbold}}_i\,\cdot\,{{\erbold}}_j }{r_j^3}   \right\}
 \label{7}
 \ee
 singled out. This disturbing potential acting on mass $\,m_i\,$ is
 generated by the masses $\,m_j\,$ other than $\,m_i\,$ or the
 primary. It deviates from the Newtonian one by the amendment
 $\;G\,m_j\,{r_{j}^{-3}}\,{{\erbold}_i\cdot{\erbold}_j }\;$
 emerging in the noninertial frame associated with the primary.

 In the simplest case of one secondary, a satellite of mass $\,m_1=M^*_{sat}\,$,
 located at a planetocentric position $\,\erbold_1\,=\,\erbold^{\;*}\,$, will be
 creating at some point $\,\erbold_2\,=\,\eRbold\;$ a perturbing potential
 \ba
 W(\eRbold\,,\;\erbold^{\;*})\;=\;-\;G\;M^*_{sat}\;\left\{\frac{1}{|\eRbold
 \;-\;\erbold^{\;*}|}\;-\;\frac{{\eRbold}\,\cdot\,{\erbold}^{\;*}}{|
 \erbold^{\;*}|^3}   \right\}\;\;\;,
 \label{8}
 \ea
 expandable over the Legendre polynomials (for $\,R\,<\,r^*\,$) by means of the formulae
 \ba
 \frac{1}{|\eRbold\;-\;\erbold^{\;*}|}\;=\;\frac{1}{r^*}\;
 \sum_{{\it{l}}=0}^{\infty}\;\left(\,\frac{R}{r^*}\,
 \right)^{\it{l}}\;P_{\it{l}} (\cos \gamma)\;\;\;
 \label{9}
 \ea
 and
 \ba
 \frac{{\eRbold}\,\cdot\,{\erbold}^{\;*}}{|
 \erbold^{\;*}|^3}\;=\;
 \frac{R\;r^{\;*}\;\cos\gamma}{{r^{\;*}}^{{\,{3}}}}\;=\;
 \frac{1}{r^{\;*}}\;\,\frac{R}{r^{\;*}}\;P_1(\cos \gamma)\;\;\;,
 \label{10}
 \ea
 $\gamma\,$ being the angular separation between $\,\eRbold\,$
 and $\,\erbold^{\;*}\,$, subtended at the point of origin, which we
 shall naturally choose to coincide with the planet's centre of mass.
 Together, the former and the latter formulae yield:
 \ba
 W(\eRbold\,,\;\erbold^{\;*})\;=\;-\;\frac{G\;M^*_{sat}}{r^{\,*}}\;
 \sum_{{\it{l}}=2}^{\infty}\,\left(\,\frac{R}{r^{\;*}}\,\right)^{
 \textstyle{^{\it{l}}}}\,P_{\it{l}}(\cos \gamma)~~~,
 \label{11}
 \label{W}
 \ea
 where we have neglected the $\,{\it{l}}=0\,$ term $~-\,GM^*_{sat}/r^{\,*}~$, because
 it bears no dependence upon $\,\eRbold\,$, and in practical problems is
 attributed to the principal part of the potential, not to the one regarded as
 perturbation. The angle $\,\gamma\,$ can be expressed via spherical coordinates as:
 \ba
 \cos\gamma\;=\;\frac{\eRbold\cdot\erbold^{\;*}}{R\;r^{\;*}}\;=\;
 \sin\phi\;\sin\phi^*\;+\;\cos\phi\;\cos\phi^*\;\cos(\lambda\,-\,\lambda^*)\;\;\;,
 \label{12}
 \ea
 $(R\,,\,\phi\,,\,\lambda)\;$ being the planetocentric distance,
 the latitude, and the longitude of the point where the disturbance
 is experienced; and $(r^*\,,\,\phi^*\,,\,\lambda^*)\;$ being the
 spherical coordinates of the satellite. It is customary (though not
 at all obligatory) to reckon the longitudes from a planet-fixed
 meridian, in which case the subsequent formulae for the potential
 come out written in a reference frame co-rotating with the planet.

 A Legendre polynomial of $\,\cos\gamma\,$, too, can be expressed via
 the spherical coordinates:
 \ba
 P_{\it{l}}(\cos \gamma)\;=\;\sum_{m=0}^{\it l}\;\frac{({\it l} - m)!
 }{({\it l} + m)!}\;(2\,-\,\delta_{0m})\;P_{{\it{l}}m}(\sin\phi)\;P_{{
 \it{l}}m}(\sin\phi^*)\;\cos m(\lambda\,-\,\lambda^*)~~~,
 \label{13}
 \ea
 substitution whereof into (\ref{W}) results in
 \ba
 W(\eRbold\,,\,\erbold^{\;*})\,=\,-\,\frac{G\;M^*_{sat}}{r^{\,*}}
 \sum_{{\it{l}}=2}^{\infty}\left(\frac{R}{r^{\;*}}\right)^{
 \textstyle{^{\it{l}}}}\sum_{m=0}^{\it l}\frac{({\it l} - m)!
 }{({\it l} + m)!}(2-\delta_{0m})P_{{\it{l}}m}(\sin\phi)P_{{
 \it{l}}m}(\sin\phi^*)\;\cos m(\lambda-\lambda^*)~~.~~~~~~
 \label{14}
 \ea
 Evidently, this formalism will stay unaltered, if the role of the tide-raising
 satellite is played by the Sun, or by another satellite, or by another planet.
 (In this case, what we call $\,M_{sat}\,$ will, in fact, denote the mass of the Sun,
 or of the other satellite, or of the other planet.)
 Likewise, the formalism may in its entirety be applied to a satellite regarded as a tidally-disturbed primary, the planet being treated as a tide-raising body (and $\,M^*_{sat}\,$ now standing for the planetary mass).

  \section{The Kaula expansion for a tidal potential}

 Kaula (1961) came up with a remarkable formula
 \ba
 \nonumber
 \left(\,\frac{1}{r^{\,*}}\,\right)^{{\it l}+1}P_{\it{l}}(\sin
 \phi^*)\;\left[\;\cos m\lambda^*\;+\;\sqrt{-1}\;\sin m\lambda^*\;\right]
 ~=~~~~~~~~~~~~~~~~~~~~~~~~~~~~~~~~~~~~~~~~~~~~~~~~~~~~~~~~~~~~~~~\\
 \label{15}
 \ea
 \ba
 \nonumber
 \left(\frac{1}{a^{\,*}}\right)^{{\it
 l}+1}\sum_{p=0}^{\infty}F_{{\it l}mp}(\inc^*)\sum_{q=\,-\,\infty}^{\infty}
 G_{{\it l}pq}(e^*)\;
  \left\{
  \begin{array}{c}
  \cos \left(\,
  v_{{\it l}mpq}^*\,-\,m\,\theta^*\,\right)\,+\,\sqrt{-1}\;\sin \left(\,
  v_{{\it l}mpq}^*\,-\,m\,\theta^*\,\right)  \\
  \sin \left(\,
  v_{{\it l}mpq}^*\,-\,m\,\theta^*\,\right)\,-\,\sqrt{-1}\;\cos \left(\,
  v_{{\it l}mpq}^*\,-\,m\,\theta^*\,\right)
  \end{array}
  \right\}^{{\it l}\,-\,m\;\;
  \mbox{\small even}}_{{\it l}\,-\,m\;\;\mbox{\small odd}~~~~~~~{\textstyle
  ,}}~~~~
  \ea
 where $F_{{\it l}mp}(\inc)\,$ are the inclination functions (Gooding and
 Wagner 2008); $\,G_{{\it l}pq}(e)\,$ are the eccentricity polynomials
 identical to the Hansen coefficients $\,X_{({\it l}-2p+q)~}^{~(\,-\,{\it
 l}-1)\,,~({\it l}-2p)}\,$; the notation $\,\sqrt{-1}\,$ is used to avoid
 confusion with the inclination; and the auxiliary combinations $\,v_{{\it l}mpq
 }^*\,$ are defined as:\footnote{~This definition agrees with that by Kaula
 (1961, 1964, 1966), but differs from the one by Lambeck (1980) who
 incorporated $\;-\,m\,\theta^*\;$ into $\,v_{{\it l}mpq}^*\,$.}
 \ba
 v_{{\it l}mpq}^*\;\equiv\;({\it l}-2p)\omega^*\,+\,
 ({\it l}-2p+q){\cal M}^*\,+\,m\,\Omega^*~~~.
 \label{16}
 \ea

 This development enabled Kaula (1961, 1964) to carry out a transformation from the
 tide-raising satellite's spherical coordinates to its orbital elements and
 the sidereal time $\,\theta^*\,$. These elements (the semimajor axis $\,a^*\,$,
 eccentricity $\,e^*\,$, inclination $\,\inc^*\,$, periapse $\,\omega^*\,$,
 ascending node $\,\Omega^*\,$, mean anomaly $\,{\cal M}^*\,$) are introduced in
 a frame that is associated with the equator but is not co-rotating with it. In
 terms of these parameters,
 \ba
 \nonumber
  W(\eRbold\,,\;\erbold^{\;*})\,=\,-\,\frac{G\,M^*_{sat}}{a^*}\sum_{{\it
  l}=2}^{\infty}\,\left(\frac{R}{a^*}\right)^{\textstyle{^{\it
  l}}}\,\sum_{m=0}^{\it l}\;\frac{({\it l} - m)!}{({\it l} + m)!}\;
  (2\,-\,\delta_{0m})\,P_{{\it{l}}m}(\sin\phi)\,\sum_{p=0}^{\it
  l}F_{{\it l}mp}(\inc^*)\,\sum_{q=\,-\,\infty}^{\infty}\,G_{{\it l}pq}
  (e^*)~~~~~~~\\
  \nonumber\\
  \label{17}\\
   \nonumber
  \left[\;\cos m\lambda\;\left\{
  \begin{array}{c}
   \cos   \\
   \sin
  \end{array}
  \right\}^{{\it l}\,-\,m\;\;
  \mbox{\small even}}_{{\it l}\,-\,m\;\;\mbox{\small odd}} \;\left(\,
  v_{{\it l}mpq}^*\,-\;m\,\theta^*\,\right)
  +\;\sin m\lambda\;\left\{
  \begin{array}{c}
  ~~\sin   \\
  -\,\cos
  \end{array}  \right\}^{{\it l}\,-\,m\;\;
  \mbox{\small even}}_{{\it l}\,-\,m\;\;\mbox{\small odd}} \;\left(\,
  v_{{\it l}mpq}^*\,-\;m\,\theta^*\,\right)
  \;\right]~~~~~~~
  \ea
 or, after carrying out the multiplication of the sine and cosine
 functions:
 \ba
 \nonumber
  W(\eRbold\,,\;\erbold^{\;*})\;=\;-\;
  \frac{G\,M^*_{sat}}{a^*}\;\sum_{{\it
  l}=2}^{\infty}\;\left(\,\frac{R}{a^*}\,\right)^{\textstyle{^{\it
  l}}}\sum_{m=0}^{\it l}\;\frac{({\it l} - m)!}{({\it l} + m)!}\;
  \left(\,2  \right. ~~~~~~~~~~~~~~~~~~~~~~~~~~~~~~~~~~~~~~~~~~~~~~~~~~~~~~~\\
                                   \nonumber\\
                                   \nonumber\\
       \left.
  ~~~ -\;\delta_{0m}\,\right)\;P_{{\it{l}}m}(\sin\phi)\;\sum_{p=0}^{\it
  l}\;F_{{\it l}mp}(\inc^*)\;\sum_{q=\,-\,\infty}^{\infty}\;G_{{\it l}pq}
  (e^*)
  \left\{
  \begin{array}{c}
   \cos   \\
   \sin
  \end{array}
  \right\}^{{\it l}\,-\,m\;\;
  \mbox{\small even}}_{{\it l}\,-\,m\;\;\mbox{\small odd}} \;\left(
  v_{{\it l}mpq}^*-m(\lambda+\theta^*)  \right)
 ~~~.~~~~~~~~~
 \label{potential}
 \label{18}
 \ea\\

 \section{Physical assumptions involved in Kaula's theory}

 If the primary is not a point mass, it becomes
 distorted by potential $\,W(\eRbold\,,\;\erbold^{\;*})\,$. The distortion of
 shape will, in its turn,
 generate some extra potential perturbation whose calculation is
 complicated by the tide-raising potential (\ref{potential}) evolving in
 time and having a rich spectrum of frequencies. The response of the primary's
 shape to each of these is different and depends on the properties of the
 planet's material. This is a situation where the linear approach becomes most
 helpful, when applicable.\footnote{~For most materials, departure from linearity
 becomes considerable when the strains approach $\,10^{-6}\,$. (Karato 2007)}

 The linear theory of bodily tides comprises two independent assertions. One is that the
 energy attenuation rate $\langle \dot{E}\rangle$ at each harmonic depends solely on
 the frequency $\,\chi\,$ and on the amplitude $\,E_{peak}(\chi)\,$, and is not influenced by
 the rest of the spectrum. This is written down as $\langle \dot{E}(\chi)\rangle=-
 \chi E_{peak}(\chi)/Q(\chi)$, which is equivalent to $\Delta E_{cycle}(\chi)=-2\pi
 E_{peak}(\chi)/Q(\chi)$, where $\Delta E_{cycle}(\chi)$ is the one-cycle energy
 loss, and $Q(\chi)$ is the quality factor. The other
 assertion is that each {\emph{stationary}} tidal change of the potential,
 $\,W_{\it l}\,$, inflicts on the planet's shape a linear deformation.
 Each of these deformations, in their turn, amend the potential of the primary
 with an addition proportional to the Love number $\,k_{\it l}\,$. As known
 from the potential theory, an addition proportional to $\,P_{\it l}(\cos\gamma)
 \,$ must be decreasing outside the spherical primary as $\,r^{-({\it l}+1)}\,
 $. Hence, were the external potential perturbation $\,W\,$ static (or,
 equivalently, were the response of the material instant), the tidal addition
 to the planetary potential would
 have assumed the form$\,$\footnote{~Following MacDonald (1964) and Singer
 (1968), we denote the tide-raising potential with $\,W\,$ and the
 bodily-tide one with $\,U\,$. In his original paper, Kaula (1964) called these
 potentials $\,U\,$ and $\,T\,$, while in the book he switched to $\,U\,$ and
 $\,U_{\textstyle{_T}}\;$
 (Kaula 1968). Be mindful that we are using a sign convention different from
 that of Kaula. As our forces are negative gradients of potentials,
 our potentials are negative to those of Kaula.}
 \ba
 \nonumber
 U(\erbold)\;=\;\sum_{{\it l}=2}^{\infty}\;k_{\it l}\;\left(\,\frac{R}{r}\,
 \right)^{{\it l}+1}\;W_{\it{l}}(\eRbold\,,\;\erbold^{\;*})~~~~~~~~~~~
 ~~~~~~~~~~~~~~~~~~~~~~~~~~~~~~~~~~~~~~~~~~~~~~~~~~~~~~~~~~~~~~\\
 \nonumber
 \ea
 \ba
 \nonumber
  =\;-\;\sum_{{\it
  l}=2}^{\infty}\;k_{\it l}\;\left(\,\frac{R}{r}\,\right)^{\textstyle{^{{\it
  l}+1}}}\frac{G\,M^*_{sat}}{a^*}\;\left(\,\frac{R}{a^*}\,\right)^{\textstyle{^{\it
  l}}}\sum_{m=0}^{\it l}\;\frac{({\it l} - m)!}{({\it l} + m)!}\;
  \left(\,2  \right. ~~~~~~~~~~~~~~~~~~~~~~~~~~~~~~~~~~~~~~~~~~~~~~~~~~~~~~~\\
                                   \nonumber\\
                                   \nonumber\\
       \left.
  ~~~ -\;\delta_{0m}\,\right)\;P_{{\it{l}}m}(\sin\phi)\;\sum_{p=0}^{\it
  l}\;F_{{\it l}mp}(\inc^*)\;\sum_{q=\,-\,\infty}^{\infty}\;G_{{\it l}pq}
  (e^*)
  \left\{
  \begin{array}{c}
   \cos   \\
   \sin
  \end{array}
  \right\}^{{\it l}\,-\,m\;\;
  \mbox{\small even}}_{{\it l}\,-\,m\;\;\mbox{\small odd}}
  \;\left(\;
  v_{{\it l}mpq}^*-m(\lambda+\theta^*) \; \right)
 ~~~.~~~~~~~~~
 \label{19}
 \ea
 $R\,$ being the mean equatorial (equivolumetric)
 radius of the planet,
 $\,\eRbold\,=\,(R\,,\,\phi\,,\,\lambda)\,$ being a particular surface point, and
 $\,\erbold\,=\,(r\,,\,\phi\,,\,\lambda)\,$ being an exterior point
 located right above the surface point $\,\eRbold\,$, at a
 planetocentric radius $\,r\,\geq\,R\,$.

 As we intend to study the effect of this potential on another external
 body, a similar transformation should be applied to the coordinates
 $\,(r\,,\,\phi\,,\,\lambda)\,$, to express $\,W\,$ through the orbital
 elements of this body. Employment of (\ref{15}), this time not for $\,
 \erbold^{\;*}\,$ but for $\,\erbold\,$, leads to:
  \pagebreak
 \ba
 \nonumber
 U(\erbold)\;=\;-\;\sum_{{\it
  l}=2}^{\infty}\;k_{\it l}\;\left(\,\frac{R}{a}\,\right)^{\textstyle{^{{\it
  l}+1}}}\frac{G\,M^*_{sat}}{a^*}\;\left(\,\frac{R}{a^*}\,\right)^{\textstyle{^{\it
  l}}}\sum_{m=0}^{\it l}\;\frac{({\it l} - m)!}{({\it l} + m)!}\;
  \left(\,2\;\right. ~~~~~~~~~~~~~~~~~~~~~~~~~~~~~~~~~~~~~~~~~~~~~~~~~~~~~\\
                                   \label{20}\\
                                   \nonumber\\
                                    \nonumber
 ~~~~~\left.-\,\delta_{0m}\,\right)\,\sum_{p=0}^{\it
  l}F_{{\it l}mp}(\inc^*)\sum_{q=-\infty}^{\infty}G_{{\it l}pq}
  (e^*)
  \sum_{h=0}^{\it l}F_{{\it
  l}mh}(\inc)\sum_{j=-\infty}^{\infty}
  G_{{\it l}hj}(e)\;\cos\left[
  \left(v_{{\it l}mpq}^*-m\theta^*\right)-
  \left(v_{{\it l}mhj}-m\theta\right) \right]
 ~~_{\textstyle{_{\textstyle ,}}}
 \ea
 a formula that generalises the tidal theory of Darwin (1908, p. 334) to
 $\,{\it l}\,$ and $\,|q|\,$ larger than $\,2\,$. Both Kaula (1964),
 who derived this milestone result, and Darwin, who had developed
 its simplified version, realised that this machinery would work only
 after the material's delayed reaction to perturbation (\ref{potential})
 is somehow taken into account. Until then (\ref{20}) remains idealised,
 in that it corresponds to an unphysical case of instantaneous response.

 To account for damping, Kaula (1964) followed the path of Darwin (1880, 1908):
 he endowed each term of the Fourier series (\ref{20}) with a real phase lag
 of its own, $\,\epsilon_{{\it l}mpq}\;$, whereafter the ultimate form of
 Kaula's expansion became
 \ba
 \nonumber
 U(\erbold)\;=\;-\;\sum_{{\it
 l}=2}^{\infty}\;k_{\it l}\;\left(\,\frac{R}{a}\,\right)^{\textstyle{^{{\it
 l}+1}}}\frac{G\,M^*_{sat}}{a^*}\;\left(\,\frac{R}{a^*}\,\right)^{\textstyle{^{\it
 l}}}\sum_{m=0}^{\it l}\;\frac{({\it l} - m)!}{({\it l} + m)!}\;
 \left(\,2\;-\right. ~~~~~~~~~~~~~~~~~~~~~~~~~~~~~~~~~~~~~~~~~~~~~~~~~\\
                                   \label{21}\\
                                   \nonumber\\
                                    \nonumber
 \left.\delta_{0m}\,\right)\,\sum_{p=0}^{\it
 l}F_{{\it l}mp}(\inc^*)\sum_{q=-\infty}^{\infty}G_{{\it l}pq}
 (e^*)
 \sum_{h=0}^{\it l}F_{{\it
 l}mh}(\inc)\sum_{j=-\infty}^{\infty}
 G_{{\it l}hj}(e)\;\cos\left[
 \left(v_{{\it l}mpq}^*-m\theta^*\right)-
 \left(v_{{\it l}mhj}-m\theta\right)-
 \epsilon_{{\it l}mpq} \right]
 ~~_{\textstyle{_{\textstyle .}}}
 \ea
 This empirical method of including dissipation into the picture contains in
 itself an important omission, of which Sir George Darwin was aware, but which
 was overlooked by his successors. Briefly speaking, even in a linear system
 a dissipation process is {\emph{not}} fully accounted for by amending phases
 of the Fourier components. This observation happens to be of relevance in the
 theory of tidal torques. We shall return to this point in section 9.

 \section{The two sidereal angles}

 Kaula's construction contains a seemingly redundant fixture, which turns out to
 be an important and useful acquisition. This is Kaula's introducing two
 sidereal angles instead of one. As these angles, $\,\theta\,$ and $\,
 \theta^*\,$, are not orbital elements of the tide-raising and tidally disturbed
 moons, but are parameters characterising the instantaneous attitude of
 the planet, it may look strange that Kaula (1964) assumed them to be different
 entities. To understand his point, let us trace the physical origin of the
 phase lag.
 The material of the primary is being deformed by a tidal stress whose spectrum
 contains an infinite number of frequencies, the reaction of the material to
 each of these being different. In a linear regime, the strain
 has the same spectrum, with each harmonic delayed by its own
 time lag $\Delta t_{\it{l}mpq}\,$. Singer (1968), and later Mignard (1979, 1980), assumed that all
 $\Delta t_{\it{l}mpq}$ are equal to one another:
 $\Delta t_{\it{l}mpq}=\Delta t$. If this were true, then in Kaula's
 series each argument $\,v_{{\it l}mpq}^*-\,m\,\theta^*$ would have to be
 substituted with
 \ba
 \nonumber
 v_{{\it l}mpq}^{*^{\;(delayed)}}-\,m\,\theta^{*^{\;(delayed)}}
 \equiv\,v_{{\it l}mpq}^*(t-\Delta t)\,-\,m\,\theta^*(t-\Delta t)
 =\,v_{{\it l}mpq}^*(t)-\,m\,\theta^*(t)\,-\,\left[\dot{v}_{{\it l}mp
 q}^*-\,m\,\dot{\theta}^*\right]\,\Delta t
 \nonumber\\
 \nonumber\\
 =\;v_{{\it l}mpq}^*(t)\;-\,m\,\theta^*(t)\,-\;\left[\;({\it l}-2p)\;\dot{\omega}^*\,+\,
 ({\it l}-2p+q)\;\dot{\cal{M}}^*\,+\,m\;(\dot{\Omega}^*\,-\,\dot{\theta}^*)\;\right]\;
 \Delta t\;\;\;.~~~~~~~~~~~~~~~~~~~~~
 \label{22}
 \ea
 In reality, however, the time lag is a function of frequency, for which
 reason the delays $\,\Delta t_{\it{l}mpq}\,$ will be different for
 each harmonic involved. This is why the arguments $\,v_{{\it l}mpq}^*
 \,-\,m\,\theta^*\,$ at the moment $\,t\,$ should rather be replaced with
 \ba
 \nonumber
 v_{{\it l}mpq}^{*^{\;(delayed)}}\;-\,m\,\theta_{{\it  l}mpq}^{*^{\;(delayed)}}
 \;=~~~~~~~~~~~~~~~~~~~~~~~~~~~~~~~~~~~~~~~~~~~~~~~~~~~~~~~~~~~~~~~~~~~~~~~~~~~
 ~~~~~
 \nonumber\\
 \label{23}\\
 \nonumber
 v_{{\it l}mpq}^*\;-\,m\,\theta^*\,-\;\left[\;({\it l}-2p)\;\dot{\omega}^*\,+\,
 ({\it l}-2p+q)\;\dot{\cal{M}}^*\,+\,m\;(\dot{\Omega}^*\,-\,\dot{\theta}^*)\;\right]\;
 \Delta t_{\it{l}mpq}\;\;\;.~~~~~~~~~~~~~~~
 \ea
 Specifically,
 \ba
 \nonumber
 v_{{\it l}mpq}^{*^{\;(delayed)}}\;=\;v_{{\it l}mpq}^*\;-\;\left[\;({\it
 l}-2p)\;\dot{\omega}^*\,+\,({\it l}-2p+q)\;\dot{\cal{M}}^*\,+\,m\;\dot{\Omega}^*\;
 \right]\;\Delta t_{\it{l}mpq}\;\;\;
 \ea
 and
 \ba
 \nonumber
 \theta_{{\it l}mpq}^{*^{\;(delayed)}}\;=\;\theta^*\;-\;
 \dot{\theta}^*\;\Delta t_{\it{l}mpq}\;\;\;,
 \ea
 $\,\dot{\theta}^*\,$ being the planet spin rate. In brief, (\ref{23}) can
 be rewritten as
 \ba
 \nonumber
 v_{{\it l}mpq}^{*^{\;(delayed)}}\,-\,m\,\theta_{{\it  l}mpq}^{*^{\;(delayed)}}
 \,=\,v_{{\it l}mpq}^*\,-\,m\,\theta^*\,-\,\omega_{\it{l}mpq}\;
 \Delta t_{\it{l}mpq}\;\;\;.~~~~~~~~~~~~~~~
 \ea
 We see that the total
 phase lags $\,\epsilon_{{\it l}mpq}\,$ introduced by Kaula are given by
 \ba
 \epsilon_{{\it l}mpq}=\left[\,({\it l}-2p)\,\dot{\omega}^*\,+\,
 ({\it l}-2p+q)\,\dot{\cal{M}}^*\,+\,m\,(\dot{\Omega}^*\,-\,\dot{\theta}^*)\,
 \right]\,\Delta t_{\it{l}mpq}=\,\omega^*_{\it{l}mpq}\,\Delta t_{\it{l}mpq}
 =\,\pm\,\chi^*_{\it{l}mpq}\,\Delta t_{\it{l}mpq}~~,~~~
 \label{24}
 \ea
 the tidal harmonic $\,\omega^*_{\it{l}mpq}\,$ being introduced as
 \ba
 \omega^*_{{\it l}mpq}\;\equiv\;({\it l}-2p)\;\dot{\omega}^*\,+\,({\it l}-
 2p+q)\;\dot{\cal{M}}^*\,+\,m\;(\dot{\Omega}^*\,-\,\dot{\theta}^*)\;~~,~~~
 \label{25}
 \ea
 the positively-defined physical frequency
 \ba
 \chi^*_{{\it l}mpq}\,\equiv\,|\,\omega^*_{{\it l}mpq}\,|\,=\,|\,({\it
 l}-2p)\,\dot{\omega}^*\,+\,({\it l}-2p+q)\,\dot{\cal{M}}^*\,+\,m\,(\dot{\Omega}^*
 \,-\,\dot{\theta}^*)\;|~~~~~
 \label{26}
 \ea
 being the actual physical $\,{{\it l}mpq}\,$ tidal frequency excited in the
 primary's material. The appropriate positively-defined time delay $\,\Delta
 t_{\it{l}mpq}\,$ depends on this physical frequency, for which reason the
 delays $\,\Delta t_{\it{l}mpq}\,$ are, generally, different from one
 another.$\,$\footnote{~When Kaula was developing his theory, the functional
 form of the dependence $\,\Delta t(\chi)\,$ was not yet known. Reliable data
 became available only in the final quarter of the past century. See formula
 (\ref{631}) below.}

 The sign on the right-hand side of (\ref{24}) is simply the sign of $\,
 \omega^*_{{\it l}mpq}\,$. The sign evidently depends on whether $\,m\,
 \dot{\theta}\,$ falls short of or exceeds the linear combination $\,({\it l}-2p)\;
 \dot{\omega}^*\,+\,({\it l}-2p+q)\;\dot{\cal{M}}^*\,+\,m\;\dot{\Omega}^*\,\approx\,
 ({\it l}-2p+q)\;\dot{\cal{M}}^*\,$.

 The origin and meaning of the phase lag $\,\epsilon_{\it{l}mpq}\,$ being now
 transparent, one may express the cosine functions in (\ref{21}) either as
 \ba
 \cos\left[\,
 \left(v_{{\it l}mpq}^*-m\theta^*\right)-
 \left(v_{{\it l}mhj}-m\theta\right)-
 \epsilon_{{\it l}mpq} \,\right]\;
 \label{27}
 \ea
 (where $\,\theta^*\,$ and $\,\theta\,$ are identical and
 cancel one another), or simply as
 \ba
 \cos\left[\,\left(v_{{\it l}mpq}^{*^{\;(delayed)}}\;-\,m\,\theta_{
 {\it l}mpq}^{*^{\;(delayed)}}\right)-
 \left(v_{{\it l}mhj}-m\theta\right)\; \right]\;\;\;.
 \label{28}
 \ea
 In (\ref{28}) we have the delayed siderial angle, $\,\theta_{{\it l}mpq}^{*^{\;
 (delayed)}}\,$, separated from the actual angle, $\,\theta\,$, by $\;-\,\dot{
 \theta}\,\Delta t_{{\it l}mpq}\,$, the time lag $\,\Delta t_{\it{l}mpq}\,$ being
 a function of $\,\chi_{{\it l}mpq}\,\equiv\,|\,\omega_{\it{l}mpq}\,|\,$.\\

 \section{The Darwin-Kaula-Goldreich expansion\\
 for the tidal torque}

 Now we are prepared to calculate the planet-perturbing tidal torque. Since in what
 follows we shall dwell on the low-inclination case, it will be sufficient to derive
 the torque's component orthogonal to the planetary equator:
 \ba
 {\tau}\;=\;-\;{M_{sat}}\;\frac{\partial U(\erbold)}{\partial\theta}\;\;\;,
 \label{29}
 \ea
 $M_{sat}\,$ being the mass of the tide-disturbed satellite, and the ``minus" sign
 emerging due to our choice not of the astronomical but of the physical sign
 convention. Adoption of the latter convention implies the emergence of
 a ``minus" sign in the expression for the potential of a point mass:
 $\;-\,GM/r\,$. This ``minus" sign then shows up on the right-hand sides of
 (\ref{6} - \ref{8}) and, later, of (\ref{19} - \ref{21}). It is then
 compensated by the ``minus" sign standing in (\ref{29}).

 The right way of calculating $\,{\partial U(\erbold)}/{\partial \theta}\,$
 is to take the derivative of (\ref{28}) with respect to $\,\theta\,$, then to
 insert (\ref{23}) into the result, and finally to get rid of the sidereal
 angle completely, by imposing the constraint $\,\theta^*\,=\,\theta\,$. This will
 yield:\footnote{~Formally, one can as well differentiate (\ref{27}) instead of
 (\ref{28}), first ignoring the fact that $\,\theta^*\,$ and $\,\theta\,$ are identical
 and then, after differentiation, permitting them to cancel one another. Though this
 method produces the same result as the rigorous calculation, it nonetheless remains a
 formal procedure lacking physics in it.}
 \ba
 \nonumber
 {\tau}= -\,\sum_{{\it
 l}=2}^{\infty}k_{\it l}\left(\frac{R}{a}\right)^{\textstyle{^{{\it
 l}+1}}}\frac{G\,M^*_{sat}\,M_{sat}}{a^*}\left(\frac{R}{a^*}\right)^{\textstyle{^{\it
 l}}}\sum_{m=0}^{\it l}\frac{({\it l} - m)!}{({\it l} + m)!}
 2m\;\sum_{p=0}^{\it
 l}F_{{\it l}mp}(\inc^*)\sum_{q=-\infty}^{\infty}G_{{\it l}pq}
 (e^*)~~~~~~~~~~~~~~~\\
                                   \nonumber\\
 \sum_{h=0}^{\it l}F_{{\it l}mh}(\inc)\sum_{j=-\infty}^{\infty} G_{{\it l}hj}(e)
 \;\sin\left[\,
 v^*_{{\it l}mpq}\,-\;v_{{\it l}mhj}\,-\;\epsilon_{{\it l}mpq}\,\right]
 ~~_{\textstyle{_{\textstyle ,}}}~~~~~~~~~~~~~~~~~~~~~~~~~~~~~~~~~~~
 \label{30}
 \ea
 In the case of the tide-raising satellite coinciding with the tide-perturbed
 one, $\,M_{sat}=M_{sat}^*\,$, and all the elements become identical to their
 counterparts with an asterisk. For a primary body not in a tidal lock with its
 satellite,\footnote{~This caveat is relevant, because in resonances expressions
 (\ref{T31} - \ref{T33}) will require modifications. For example, the sidereal angle
 of a satellite tidally locked in a $\,1:1\,$ resonance will be: $\;\,\theta\,=
 \,\Omega\,+\,\omega\,+\,{\cal M}\,+\,180^o\,+\,\alpha\,+\,O(i^2)\;\,$, letter
 $\alpha\,$ denoting the librating angle, which is subject to damping and
 therefore is normally small (less than $\,2"\,$ for the Moon). Inserting the
 said formula for $\,\theta\,$ into the expression (\ref{25}) for the tidal
 harmonic, we obtain, in neglect of $\,\;-m\dot{\alpha}\,$:
 \ba
 \nonumber
 \omega^*_{{\it{l}}mpq}\;\equiv\;({\it l}-2p-m)\;\dot{\omega}^*\,+\,({\it l}-
 2p+q-m)\;\dot{\cal{M}}^*\;\;~~.~~~
 \ea
 We now see that, since $\,\theta\,$ is a function of the other angles,
 different sets of the indices's values will correspond to one value of the
 tidal frequency. We shall illustrate this by considering the so-called
 anomalistic modes $\,\pm\dot{{\cal{M}}}\,$ in the potential. These modes,
 corresponding to the physical frequency $\,|\dot{{\cal{M}}}|\,$, are given by
 $\,({\it{l}}mpq)\,=\,(201,\pm 1)~$ and also by $\,({\it{l}}mpq)\,=\,(220,\pm1)
 \,$. Although the $\,m=0\,$ terms enter the potential, they will not be in the
 torque, as can be observed by differentiating (\ref{19}) with respect to $\,
 \lambda\,$, or by differentiating equation (\ref{21}) with respect to $\;-
 \theta\;$, or simply by noticing the presence of the factor $\,m\,$ on the
 right-hand side of (\ref{30}). Nonetheless, we see that there exists a pair of
 $\,m=2\,$ terms, which provides an anomalistic input into the torque. This way,
 the case of libration deserves a separate consideration, as it is more involved
 than that of tidal despinning. Specifically, in the case of libration a value
 of the tidal frequency may correspond to different sets of the indices' values.} the torque (\ref{30}) can be split into two
 parts. The first part is constituted by those terms of (\ref{30}), in which
 indices $\,(p\,,\,q)\,$ coincide with $\,(h\,,\,j)\,$, and therefore all
 $\,v_{{\it{l}}mhj}\,$ cancel with $\,v_{{\it l}mpq}^*\,$, provided the
 tidally-perturbed satellite and the tide-raising one are the same body. This
 component of the torque is, therefore, constant. The rest of the total sum
 (\ref{30}) will be denoted with $\,\tilde{\tau}\,$. It is comprised of the
 terms, in which the pairs $\,(p\,,\,q)\,$ differ from $\,(h\,,\,j)\,$.
 Accordingly, these terms contain the differences
 \ba
 \nonumber
 v^*_{{\it{l}}mpq}\,-\;v_{{\it{l}}mhj}\,=~~~~~~~~~~~~~~~~
 ~~~~~~~~~~~~~~~~~
 \ea
 \ba
 ({\it{l}}\,-\,2\,p\,+\,q)\;{\cal{M}}^*\,-\;({\it{l}}\,-\,2\,h\,+\,j)\,{\cal{M}}\;+\;m\,( {\Omega}^*\, - {\Omega} )\;+\;{\it{l}}\;( {\omega}^*\,-\,{\omega} )\,-\,2\,p\,{\omega}^*
 \,+\,2\,h\,{\omega}~~~.~~~
 \label{}
 \ea
 When the tidally-perturbed and tide-raising moons are the same body, this becomes
 \ba
 v^*_{{\it{l}}mpq}\,-\;v_{{\it{l}}mhj}\,=~(2\,h\,-\,2\,p\,+\,q\,-\,j)\;{\cal{M}}\,+
 \,(2\,h\,-\,2\,p)\,{\omega}~~~,~~~
 \label{bobo}
 \ea
 whence we see that the oscillating component of the torque, $\,\tilde{\tau}\,$, consists
 of two parts. The part with $\;h\,-\,p\,=\,0\;$ and $\,q\,-\,j\neq\,0\;\,$ consists solely
 of short-period terms, and it averages out trivially.

 The mixed-period part of (\ref{30}), with $\,h\,-\,p\,\neq\,0\;$, consists of both a short-period contribution dependent upon the mean anomaly, and a long-period contribution depending upon the argument of pericentre. All such terms contain multipliers like $\;F_{2mp}(\inc)\,F_{2mh}(\inc)\;$, where $\,h\,\neq\,p\;$, and $\;F_{220}\,=\,3\,+\,O({\inc}^2)\;$, $\;\,F_{210}\,=\,3/2\,\sin
 \inc\,+\,O({\inc}^3)\;$, $\;\,F_{211}\,=\,-\,3/2\,\sin\inc\,+\,O(\inc^3)\;$,
 $\;\,F_{221}\,=\,3/2\,\sin^2\inc\;$, the other relevant $\,F_{2mn}$'s being of higher order than $\,O({\inc}^2)\,$. So the only long-period terms that we have to consider in (\ref{30}) involve products: $\;F_{210}(\inc)\,F_{211}(\inc)\;$, $\,\;F_{211}(\inc)\,F_{210}(\inc)\;$, $\;F_{220}(\inc)\,F_{221}(\inc)\;$, and $\,\;F_{221}(\inc)\,F_{220}(\inc)\;$. However these products are of order $\,O({\inc}^2)\,$. Thus, while the short-period terms in (\ref{30}) average out over one rotation period of the moon about the planet, the long-period terms are of order  $\,O(e^2\inc^2)\,$, the $\,e^2\,$ coming from the $\,G_{2pq}\,$ functions. (Indeed, when $\,h\,$ and $\,p\,$ differ by $\,1\,$, then $\,q\,$ and $\,j\,$ must differ by $\,2\,$, to eliminate the mean anomaly, i.e., to make the term long-period and not short-period.)
 So both the short- and long-period contributions may be neglected in our approximation.\footnote{~Had we tried to expand our treatment to higher
 inclinations, our neglect of
 the short-period terms would remain legitimate, for they still would average out over
 one rotation period of the satellite about its primary.

 As for the long-period terms, it would be tempting to say that these average out over
 the apsidal-precession period. The latter is much shorter than the time scale of the
 planetary spin deceleration, a circumstance that may seem a safe justification for the
 neglect of the long-period terms also for higher inclinations. However, a word of
 warning would be appropriate here. As well known from Kozai (1959a), who took into account
 the primary's nonsphericity, the pericentre of a satellite inclined by about
 $\,63^o\,$ or $\,117^o\,$ will neither advance nor retard, at least within the
 first-order (in $\,J_2\,$) perturbation theory. (For a critical review of
 Kozai's theory see Taff 1985.) Kozai's original attempt to introduce corrections
 owing to $\,J_3\,$ and $\,J_4\,$ was flawed because in the vicinity of the
 critical inclinations these terms should be considered not as higher-order but
 rather as leading. His later analysis demonstrated that at these inclinations
 the satellite's perigee should librate about $\,90^o\,$ or $\,270^o\,$ (Kozai
 1962). Under these circumstances, the long-period terms in our expression for
 the torque will {\underline{not}} be averaged out. We however may neglect this
 possibility, because in the current work we consider only low-inclined moons.

 Another situation, which we exclude from our treatment, is libration of the
 satellite's periapse about $\,90^o\,$ or $\,270^o\,$, caused by the pull of
 a third body (the star or some large neighbouring planet). The possibility
 of such librations may be derived from the presence of the $\,\cos2\omega\,$
 term on the right-hand side of the equation for $\,d\omega/dt\,$ in the
 theory of Kozai (1959b, 1962) -- for an easy introduction into this theory
 see Innanen et al. (1997), and for its generalisation to finite obliquities
 see Gurfil et al. (2007).
 An important special case of the theory of the third-body-caused librations
 is the one of
 the satellite getting into a resonance with the third body. An indication that
 such resonances may cause the satellite's $\,\omega\,$ librate comes from the
 mathematically similar theory of Pluto-Neptune resonances (Williams \& Benson
 1971): being in resonance with Neptune, Pluto has its periapse librating due
 to a high inclination. (To be exact, the behaviour of Pluto's periapse is
 dictated not only by Neptune, but by the combined influence of all of the four
 gas giant planets. However, this does not change the main point: the outer body
 or bodies can cause apsidal libration.)

 In the Solar system, none of the large satellites is so highly perturbed as to
 have a periapse librating around $\,90^o\,$ or $\,270^o\,$ due to the above two
 mechanisms. In theory, though, this remains an option for exoplanets. Either
 librating mechanism might apply also to satellites of minor planets. In our
 current paper we do not consider such moons.}

 Thus we arrive at:
 \ba
 {\tau}~=
 \sum_{{\it{l}}=2}^{\infty}2~k_{\it l }~G~M_{sat}^{\textstyle{^{2}}}~
 \frac{R^{\textstyle{^{2{\it{l}}\,+\,1}}}}{
 a^{\textstyle{^{2\,{\it{l}}\,+\,2}}}}
 \sum_{m=0}^{\it l}
 \frac{({\it{l}}\,-\,m)!}{({\it{l}}\,+\,m)!}
 \;m\;\sum^{\it l}_{p=0}
 \;F^{\textstyle{^{2}}}_{{\it{l}}mp}(\inc)\sum^{\it \infty}_{q=-\infty}G^{\textstyle{^{2}}}_{{\it{l}}pq}(e)
 \;\sin\epsilon_{{\it{l}}mpq}\;+\;\tilde{\tau}\;\;\;,~~~
 \label{31}
 \label{T31}
 \ea
 the sum standing for the constant (${\cal{M}}$-independent) part of the torque,
 and $\tilde{\tau}$ denoting the oscillating part whose time-average is zero.

 As we pointed in the end of section 5, the sign of the phase lag $\,
 \epsilon_{{\it{l}}mpq}\,$ depends on whether $\,m\,\dot{\theta}\,$ falls short
 of or exceeds the linear combination $\,({\it{l}}-2p)\;\dot{\omega}^*\,+\,(
 {\it{l}}-2p+q)\;\dot{\cal{M}}^*\,+\,m\;\dot{\Omega}^*\,\approx\,({\it{l}}-2p
 +q)\;\dot{\cal{M}}^*\,$. Now we also understand that, outside resonances,
 the $\,{{\it{l}}mpq}\,$ component of the tidal
 torque experienced by the planet is decelerating if the values of $\,m\,
 \dot{\theta}\,$ exceed the given combination, and is accelerating otherwise.

 Expression (\ref{31}) gets considerably simplified if we restrict ourselves to
 the case of $\,{\it l}\,=\,2\,$. Since $\,0\,\leq\,m\,\leq\,{\it l}\,$,
 and since $\,m\,$ enters the expansion as a multiplier, we see that only
 $\,m\,=\,1\,,\,2\,$ actually matter. As $\,0\,\leq\,p\,\leq\,{\it l}\,$,
 we are left with only six relevant $\,F$'s, those corresponding to
 $\;(\it{l}mp)\,=\,$ (210), (211), (212), (220), (221), and (222). By a
 direct inspection of the table of $\,F_{\it{l}mp}\,$ we find that five
 of these six functions happen to be $\,O(\inc)\,$ or $\,O(\,\inc^2\,)\,$,
 the sixth one being $\,F_{220}\,=\,\frac{\textstyle 3}{\textstyle
 4}\,\left(\,1\,+\,\cos\inc\,\right)^2\,=\,3\,+\,O(\inc^2)\,$. Thus, in the
 leading order of $\,\inc\;$, the constant part of the torque reads:
 \ba
 {\tau}_{\textstyle{_{\textstyle_{\textstyle{_{l=2}}}}}}~=~\frac{3}{2}
 ~\sum_{q=-\infty}^{\infty}~G~M_{sat}^{\textstyle{^{2}}}~~R^{\textstyle{^{5}}}\;
 a^{-6}\;G^{\textstyle{^{2}}}_{\textstyle{_{\textstyle{_{20\mbox{\it{q}}}}}}}
 (e)\;k_{{2}}\;\sin\epsilon_{\textstyle{_{\textstyle{_{220\mbox{\it{q}}}}}}}
 \;+\;O(\inc^2/Q)\;\;\;.
 \label{32}
 \ea
 This is what is called Darwin-Kaula-Goldreich torque, or simply Darwin
 torque. The principal term of this series is
 \ba
 {\tau}_{\textstyle{_{\textstyle{_{\textstyle{_{2200}}}}}}}~=~\frac{3}{2}
 ~G\,M_{sat}^2~k_2~R^{\textstyle{^5}}\;a^{{{-\,6}}}
 \sin\epsilon_{\textstyle{_{2200}}}\;\;\;.
 \label{T33}
 \label{33}
 \ea
 Switching from the lags to quality factors via formula\footnote{~The phase
 lag $\,\epsilon_{\it{l}mpq}\,$ is introduced in (\ref{23} - \ref{24}), while
 the tidal harmonic $\,\omega_{\it{l}mpq}\,$ is given by (\ref{25}). The
 quality factor $\,Q_{\it{l}mpq}\,=\,|\,\cot \epsilon_{\it{l}mpq}\,|$ is, for
 physical reasons, positively defined. Hence the multiplier
 $\,\mbox{sgn}\,\omega_{\it{l}mpq}\,$ in (\ref{34}). (As ever, the
 function $\,\mbox{sgn}(x)\,$ is defined to assume the values $\,+1\,$, $\,-1\,$,
 or $\,0\,$ for positive, negative, or vanishing $\,x\,$, correspondingly.)

 Mind that no factor of two appears in (\ref{Q} - \ref{epsilon}), because
 $\epsilon$ is a phase lag, not a geometric angle.}
 \ba
 Q_{\it{l}mpq}\,=\,|\,\cot \epsilon_{\it{l}mpq}\,|\;\;\;,
 \label{Q}
 \ea
 we obtain:
 \ba
 \sin\epsilon_{\textstyle{_{\textstyle{_{{{{\it{l}}mpq}}}}}}}=\,
 \sin|\epsilon_{\textstyle{_{\textstyle{_{{\it{l}}mpq}}}}}|\;\,\mbox{sgn}\,
 \omega_{\textstyle{_{\textstyle{_{{\it{l}}mpq}}}}}=\,\frac{\mbox{sgn}\,
 \omega_{\textstyle{_{\textstyle{_{{\it{l}}mpq}}}}}\;}{\sqrt{{\textstyle 1~+~\cot^2
 \epsilon_{\textstyle{_{\textstyle{_{{\it{l}}mpq}}}}}}}}=\;\frac{\mbox{sgn}\,
 \omega_{\textstyle{_{\textstyle{_{{\it{l}}mpq}}}}}\;}{\sqrt{{\textstyle 1~+~
 Q^{\textstyle{^{2}}}_{\textstyle{_{\textstyle{_{{\it{l}}mpq}}}}}}}}
 =~\frac{~\mbox{sgn}\,
 \omega_{\textstyle{_{\textstyle{_{{\it{l}}mpq}}}}}~}{Q_{
 \textstyle{_{\textstyle{_{{\it{l}}mpq}}}}}}+O(Q^{-3})~~,~~~
 \label{34}
 \label{epsilon}
 \ea
 whence
 \ba
 \nonumber
 {\tau}_{\textstyle{_{\textstyle_{\textstyle{_{l=2}}}}}}~=~
 \frac{3}{2}~\sum_{q=-\infty}^{\infty}~G~M_{sat}^2~\;R^{\textstyle{^{5}}}\;a^{-6}
 \;G^{\textstyle{^{2}}}_{\textstyle{_{20\mbox{\it{q}}}}}(e)\;
 k_{\textstyle{_2}}\;\frac{~\mbox{sgn}\,
 \omega_{\textstyle{_{220\mbox{\it{q}}}}}\,}{
 Q_{\textstyle{_{\textstyle{_{220\mbox{\it{q}}}}}}}}\;\,
 \,+\,O(\inc^2/Q)\,+\,O(Q^{-3})\;\;\;.
 \label{}
 \ea
 Now, let us simplify the sign multiplier. If in expression (\ref{25}) for
 $\omega_{\textstyle{_{{\it{l}}mpq}}}$ we get rid of the
 redundant asterisks, replace\footnote{~While in the undisturbed two-body
 setting $\,{\cal{M}}\,=\,{\cal{M}}_0+n\,(t-t_0)\,$ and $\,\dot{\cal{M}}
 =n\,$, under perturbation these relations get altered. One possibility is to
 introduce (following Tisserand 1893) an {\emph{osculating mean motion}}
 $\,n(t)\,\equiv\,\sqrt{\mu/a(t)^3}\,$, and to stick to this definition under
 perturbation. Then the mean anomaly will evolve as $\,\;{\cal{M}}\;=\;{\cal{M}}_{\textstyle{_0}} (t)
 \,+\;\int_{\textstyle{_{t_{_0}}}}n(t)\;dt\;$, whence $\;\dot{\cal{M}} =\dot{\cal{M}}_{\textstyle{_0}} (t)+n(t)\,$.

 Other possibilities include introducing an {\emph{apparent}} mean motion, i.e.,
 defining $\,n\,$ either as the mean-anomaly rate $\,d{\cal M}/dt\,$, or as the
 mean-longitude rate $\,dL/dt\,=\,d\Omega/dt\,+\,d\omega/dt\,+\,d{\cal{M}}/dt\,$
 (as was done by Williams et al. 2001). It should be mentioned in this regard that,
 while the first-order perturbations in $a(t)$ and in the osculating mean motion
 $\sqrt{\mu/a(t)^3}$ do not have constant parts leading to secular rates, the epoch terms typically do have
 secular rates. These considerations explain why there exists a difference between the apparent mean motion defined as
 $dL/dt$ (or as $d{\cal M}/dt$) and the osculating mean motion $\sqrt{\mu/a(t)^3}$
 .

 In many practical situations, the secular rate in $\,{\cal{M}}_0\,$ is of the order
 of the periapse rate, while the secular rate in $\,L_0\,$ turns out to be smaller.
 Hence the advantage of defining the apparent $\,n\,$ as the mean-longitude rate
 $\,dL/dt\,$, rather than as the mean-anomaly rate $\,d{\cal{M}}/dt\,$. (At the same time,
 for a satellite orbiting an oblate planet the secular rates of $\,M_0\,$, $\,L_0\,$,
 and periapse are of the same order.)

 Although the causes of orbit perturbations are beyond the scope of our paper,
 we would mention that in the expression (\ref{25}) for
 $\omega_{\textstyle{_{{\it{l}}mpq}}}$ the notations $\,\dot{\cal{M}}\,$,$\,\dot{\omega}\,$,
 and $\,\dot{\Omega}\,$ generally imply the {\emph{secular}} rate.} $\;\dot{\cal{M}}$ with $\dot{\cal{M}}_0+n\approx n$, and set ${\it{l}}=m=2$
 and $p=0$, the outcome will be:
 \ba
 \nonumber
 \mbox{sgn}\,\omega_{\textstyle{_{220\mbox{\it{q}}}}}\;=\;
 \mbox{sgn}\,\left[\,2\;\dot{\omega}\,+\,(2+q)\;n\,+\,2\,\dot{\Omega}-\,2\,\dot{\theta}
 \,\right]\;=\;\mbox{sgn}\,\left[\,\dot{\omega}\,+\,\left(1\,+\,
 \frac{\textstyle q}{\textstyle 2}\,\right)\;n\,+\,\dot{\Omega}-\,\dot{\theta}
 \,\right]~~~.
 \label{}
 \ea
 As the node and periapse precessions are slow,
 the above expression may be simplified to
 \ba
 \nonumber
 \mbox{sgn}\,\left[\,\left(1\,+\,
 \frac{\textstyle q}{\textstyle 2}\,\right)\;n\,-\,\dot{\theta}
 \,\right]~~~.
 \label{}
 \ea
 All in all, the approximation for the constant part of the torque assumes the form:
  \ba
 {\tau}_{\textstyle{_{\textstyle_{\textstyle{_{l=2}}}}}}\,=\,
 \frac{3}{2}~\sum_{q=-\infty}^{\infty}~G~M_{sat}^2~\;R^{\textstyle{^{5}}}\;a^{-6}
 \,G^{\textstyle{^{2}}}_{\textstyle{_{20\mbox{\it{q}}}}}(e)\;
 k_{\textstyle{_2}}\;{Q^{\textstyle{^{-1}}}_{\textstyle{_{\textstyle{_{220\mbox{\it{q}}}}}}}}
 \;~\mbox{sgn}\,\left[\,\left(1\,+\,
 \frac{\textstyle q}{\textstyle 2}\,\right)\;n\,-\,\dot{\theta}
 \,\right]
 +O(\inc^2/Q)+O(Q^{-3})\;\;.~~~
 \label{35}
 \label{exp}
 \ea
 That the sign of the right-hand side in the above formula is correct can be
 checked through the following obvious observation: for a sufficiently high
 spin rate $\,\dot{\theta}\,$ of the planet, the multiplier $\,\mbox{sgn}\,
 \left[\,\left(1\,+\,\frac{\textstyle q}{\textstyle 2}\,\right)\;n\,-\,
 \dot{\theta}\,\right]\,$ becomes negative. Thereby the overall expression
 for $\,{\tau}_{\textstyle{_{\textstyle_{\textstyle{_{l=2}}}}}}\,$
 acquires a ``minus" sign, so that the torque points out in the direction
 of rotation opposite to the direction of increase of the sidereal
 angle $\,\theta\,$. This is exactly how it should be, because for a
 fixed $\,q\,$ and a sufficiently fast spin the $\,q$'s component of the
 tidal torque must be decelerating and driving the planet to synchronous
 rotation.

 Expansion (\ref{exp}) was written down for the first time, without proof, by
 Goldreich \& Peale (1966). A schematic proof was later offered by Dobrovolskis (2007).
 %

 \section{The MacDonald expression for the tidal torque}

 The idea of representing the tidal pattern with one bulge belongs to MacDonald (1964). Later, Singer (1968) and  Mignard (1979, 1980) realised that MacDonald's single-bulge simplification was acceptable only with a frequency-independent $\,\Delta t\,$, not with a frequency-independent $\,Q\,$
 as in MacDonald (1964). Nevertheless we shall call this approach ``the MacDonald torque",
 to comply with the established convention. For the same reason, the afore-described
 Darwin-Kaula-Goldreich expansion will be referred to simply as ``the Darwin torque".

 In the preceding section, the Darwin torque's component orthogonal to the equator
 was conveniently given by the fundamental formula (\ref{29}).
  Within the MacDonald approach,
  it will be more practical to write the torque as a derivative taken with respect to the
  longitude. The torque acting on the tidally disturbed satellite of mass $\,M_{sat}\,$ is
  $\;-\,M_{sat}\;\partial U/\partial \lambda$, while the torque that this moon exerts on
  the planet is this expression's negative:
  \ba
  \tau(\erbold)\;=\;M_{sat}\;\frac{\partial U(\erbold)}{\partial\lambda}~~~.~~~
  \label{415}
  \ea
  Speaking rigorously, the formula furnishes the torque's component perpendicular
  to the planetary equator. As can be seen from (\ref{38}), formula (\ref{415})
  coincides with (\ref{29}) for low inclinations.

 \subsection{Simplifications available for low \it{\bf{\emph{i}}}}

 In principle, we can as well insert into
 \ba
 \nonumber
 U(\erbold)\;=\;\sum_{{\it l}=2}^{\infty}\;k_{\it l}\;\left(\,\frac{R}{r}\,
 \right)^{{\it l}+1}\;W_{\it{l}}({\eRbold}\,,\;\erbold^{\;*})~~~~~~~~~~~
 ~~~~~~~~~~~~~~~~~~~~~~~~~~~~~~~~~~~~~~~~~~~~~~~~~~~~~~~~~~~~~~
 \label{}
 \ea
 the ``raw" expression (\ref{14}), the one as yet ``unprocessed" by (\ref{15}).
 This will give us
 \ba
 U(\erbold)\;=\;\,-\,{G\;M_{sat}^*}
 \sum_{{\it{l}}=2}^{\infty}k_{\it l}\;
 \frac{R^{
 \textstyle{^{2\it{l}+1}}}}{r^{
 \textstyle{^{\it{l}+1}}}{r^{\;*}}^{
 \textstyle{^{\it{l}+1}}}}\sum_{m=0}^{\it l}\frac{({\it l} - m)!
 }{({\it l} + m)!}(2-\delta_{0m})P_{{\it{l}}m}(\sin\phi)P_{{
 \it{l}}m}(\sin\phi^*)\;\cos m(\lambda-\lambda^*)~~~~~~~~
 \label{36}
 \ea
 or, for low inclinations of both the tidally-perturbed and tide-raising satellites:
 \ba
  \nonumber
 U(\erbold)=-{GM_{sat}^*}
 \sum_{{\it{l}}=2}^{\infty}k_{\it l}
 \frac{R^{
 \textstyle{^{2\it{l}+1}}}}{r^{
 \textstyle{^{\it{l}+1}}}{r^{\;*}}^{
 \textstyle{^{\it{l}+1}}}}
 \sum_{m=0}^{\it l}\frac{({\it l} - m)!
 }{({\it l} + m)!}(2-\delta_{0m})P_{{\it{l}}m}(0)P_{{
 \it{l}}m}(0)\cos m(\lambda-\lambda^*)~~~~~~~~~~~~~~~~~~~~~~~~~\\
 \nonumber\\
 \left.~\right.~~~~~~~~~~~~~~~~~~~~
 ~~~~~~~~~~~~~~~~~~~~~~~~~~~~~~~~~~~~
 ~~~~~~~~~~~~~~~~~~~~~~~~~~~~~~~~+O(\inc^2)+O({\inc^*}^{2})+O(\inc\inc^*)~~.~~~
 \label{4437}
 \ea
 At this point we once again are faced with the question of how to
 bring damping into the picture, i.e., how to take care of the delayed
 reaction of the planet's material to the tidal stress. It is
 tempting to substitute $m\lambda^{\textstyle^*}$ with its delayed value.
 Then instead of $\cos m (\lambda-\lambda^*)$ we get
 \ba
 \cos\left(~m~\lambda~-~m~\lambda^{\textstyle^{*^{\,\textstyle{^{(delayed)}}}}}~\right)
 ~=~\cos \left(\;m\;\lambda\;-\;\left[\,m\lambda^{\textstyle^*
 }\,-\,m\stackrel{\centerdot}{\lambda}{^{\textstyle^*}}\Delta t\,\right]\;\right)
 \;\;\;,
 \label{trick}
 \ea
 This trick, suggested by Kaula (1968, page 201),\footnote{~Mind the difference in
 notations. While in the original paper Kaula (1964) denoted the phase lags with
 $\,\epsilon_{\textstyle{_{\it{l}mpq}}}\,$, in his book Kaula (1968) called them
 $\,\varphi_{\textstyle{_{\it{l}mpq}}}\,$. For the longitudinal lag $\,2
 \stackrel{\centerdot}{\lambda}{^{\textstyle^*}}\Delta t\,$ emerging in our formula
 (\ref{40}), Kaula (1968) used notation $\,2\delta\,$. This way, in the terms used
 by Kaula (1968) in his book, the geometric angle subtended at the primary's centre
 between the directions to the bulge and the moon is called $\,2\,\delta\,$, not $\,
 \delta\,$ as in most literature.} has a physical justification only if $\,\Delta t
 \,$ is the same for all frequencies, a model pioneered by Singer (1968) and
 furthered by Mignard (1979, 1980). It can be shown that this model is equivalent to
 the following rheological law:\footnote{~Combining (\ref{24}) with the relation
 $\,Q=1/\tan|\epsilon |\,$, we see that setting all
 $\,\Delta t_{\textstyle{_{\it{l}mpq}}}\,$ equal to the same $\,\Delta t\,$ is
 equivalent to saying that the quality factor scales as the inverse frequency:
 $\;Q\,=\,{1}/({\chi\;\Delta t})\;$, provided, of course, that the $Q$ factor is large.
 As can be seen from (\ref{34}), a more exact relation will read:
 $\;\sin(\chi\,\Delta t)\;=\;{1}/{\sqrt{1\,+\,Q^2}}\;$,
 so that $\,\chi\,\Delta t\,=\,Q^{-1}\,+\,O(Q^{-3})\,$.

 Very special is the case when the values of the quality factor are very low (say, much less than 10). In this situation,
the interconnection between the quality factor and the phase lag becomes quite
different from the customary formula $\,Q\,=\,\cot\epsilon\,$ . See the
Appendix for details.}
 \ba
 Q_{\textstyle{_{\it{l}mpq}}}\;=\;\frac{1}{\chi_{\textstyle{_{\it{l}mpq}}}\,\;\Delta t}\;\;\;.
 \label{model}
 \ea
 Even then, though, it
 remains unclear how to connect the longitude lag $\,m
 \stackrel{\centerdot}{\lambda}{^{\textstyle^*}}\Delta t\,\,$ with
 one or another $\,Q_{\textstyle{_{\it{l}mpq}}}\,$, in the spirit of (\ref{34}).
 To see what can be done, write down the longitude (reckoned from a fixed
 meridian on the rotating planet) as
 \ba
 \lambda\;=\;-\;\theta\;+\;\Omega\;+\;\omega\;+\;\nu\;+\;O(\inc^2)\;=\;
 -\;\theta\;+\;\Omega\;+\;\omega\;+\;{\cal{M}}\;+\;2\;e\;\sin{\cal{M}}\;+
 \;O(e^2)\;+\;O(\inc^2)\;\;\;,\;\;\;\;
 \label{38}
 \ea
 $\nu\,$ being the true anomaly. Thence, in neglect of the nodal and apsidal
 precessions, the cosine becomes:
 \ba
 \cos \left(\;\left[\,m\;\lambda\;-\;m\lambda^{\textstyle^*}\,\right]\;+\,
 m\stackrel{\centerdot}{\lambda}{^{\textstyle^*}}\Delta t\;\right)\;=\;
 \cos \left(\;\left[\,m\;\lambda\;-\;m\lambda^{\textstyle^*}\,\right]\;+\,
 m\;\left[\dot{\nu}^*\,-\;\dot{\theta}^*\right]\;\Delta t\;\right)\;\;\;,\;\;\;\;\;
 \label{39}
 \ea
 or, equivalently:
 \ba
 \cos \left(\left[m\lambda-m\lambda^{\textstyle^*}\right]+
 m\stackrel{\centerdot}{\lambda}{^{\textstyle^*}}\Delta t\right)=
 \cos \left(\left[m\lambda-m\lambda^{\textstyle^*}\right]+
 m\left[n^*-\dot{\theta}^*\right]\Delta t+2me^*n^*\Delta t\,\cos{\cal M}^*
 +O(e^2)\right)\;\;\;.\;\;\;\;
 \label{40}
 \label{4446}
 \ea
 Insertion of (\ref{39}) into (\ref{4437}), along with substitution of $\,r^*(t)\,$
 by $\,r^*(t-\Delta t)\,$, leads us to
 \ba
 \nonumber
 U(\erbold)=-{GM_{sat}^*}
 \sum_{{\it{l}}=2}^{\infty}k_{\it l}
 \frac{R^{
 \textstyle{^{2\it{l}+1}}}}{r(t)^{
 \textstyle{^{\it{l}+1}}}{r^{^*}}(t-\Delta t)^{
 \textstyle{^{\it{l}+1}}}}
 \sum_{m=0}^{\it l}\frac{({\it l} - m)!
 }{({\it l} + m)!}(2-\delta_{0m})P_{{\it{l}}m}(0)P_{{
 \it{l}}m}(0)\,\cos
 \left(\;m\,\left[\,\lambda-\lambda^{\textstyle^*}\,\right]  \right.~~~~~~~\\
 \label{837}
 \label{G837}\\
 \nonumber
 \left. +\;
 m\,\left[\dot{\nu}^*-\dot{\theta}^*\right]\,\Delta t\;\right)
 +O(\inc^2)+O({\inc^*}^{2})+O(\inc\inc^*)~~.~~~
 \label{837}
 \ea
 If we take into account only the $\;{\it l}\,=\,2\;$ contribution,
 expression (\ref{4437}) will simplify to
 \ba
 U(\erbold)=\;-\;\frac{{G\;M_{sat}^*}\,k_{2}\;R^{
 \textstyle{^5}}}{r(t)^{
 \textstyle{^{3}}}{r^{^*}}(t)^{
 \textstyle{^{3}}}}
 \sum_{m=0}^{2}\frac{(2 - m)!
 }{(2 + m)!}(2-\delta_{0m})P_{2m}(0)P_{{
 2}m}(0)\,\cos m(\lambda-\lambda^*)+O(\inc^2)+O({\inc^*}^{2})+O(\inc\inc^*)~~,~~
 \label{937}
 \ea
 where only the $\,m\,=\,2\,$ term is important.\footnote{~In (\ref{937}), we
 may neglect the $\,\lambda$-independent term with $\,m\,=\,0\,$, because
 our eventual intention is to find the torque by differentiating $\,U(
 \erbold)\,$ with respect to $\,\lambda\,$. We may also omit the $\,m\,=\,1
 \,$ term, because $\;P_{21}(0)\,=\,0\;$. This omission brings up an error of
 order $\,O(\inc\inc^*)\,$ into equations (\ref{4437}), (\ref{G837} - \ref{42}), and (\ref{44})} In the presence of dissipation,
 the appropriately simplified version of (\ref{937}) will read:
 \ba
 U(\erbold)=\,-\,\frac{3}{4}\;
 \frac{G\;M_{sat}^*\;k_{2}\;R^{\textstyle{^5}}}{r(t)^{\textstyle{^{3}}}{
 r^{^*}}(t-\Delta t)^{\textstyle{^{3}}}} \;\cos \left(\;\left[\,2\;\lambda\;-\;2\lambda^{\textstyle^*}\,\right]\;+\,
 2\;\left[\dot{\nu}^*\,-\;\dot{\theta}^*\right]\;\Delta
 t\;\right)\;+\;O(\inc^2)+O({\inc^*}^2)+O(\inc\inc^*)~~,~~~
 \label{4441}
 \ea
 while the corresponding expression for the torque exerted by the satellite on
 the planet will, in this approximation, be given by
 \ba
 \tau(\erbold)=M_{sat}\frac{\partial U(\erbold)}{\partial\lambda}=
 \frac{3\,GM_{sat}^*M_{sat}k_{2}R^{\textstyle{^5}}}{2\,r(t)^{\textstyle{^{3}}}{
 r^{^*}}(t-\Delta t)^{\textstyle{^{3}}}} \;\sin \left(\left[2\lambda-2\lambda^{\textstyle^*}\right]
 +2\left[\dot{\nu}^*-\dot{\theta}^*\right]\Delta t\right)+O(\inc^2)+O({\inc^*}^2)+O(\inc\inc^*)~~.~~
 \label{42}
 \ea
 In the case when the tidally disturbed satellite coincides with the tide-raising
 one, i.e., when $\,\lambda\,=\,\lambda^*\,$ and $\,M_{sat}\,=\,M_{sat}^*\,$, we obtain:
 \ba
 \nonumber
 \tau &=& \frac{3}{2}\,{G\,M_{sat}^2}\;k_{2}\;
 \frac{R^{\textstyle{^5}}}{r(t)^{\textstyle{^{3}}}{
 r}(t-\Delta t)^{\textstyle{^{3}}}} \,\sin
 \left(2\left[\dot{\nu}-\,\dot{\theta}\right]\Delta t\right)
 +O(\inc^2/Q)\\
 \label{43}\\
 \nonumber
 &=&\frac{3}{2}\;{G\,M_{sat}^2}\;k_{2}\;
 \frac{R^{\textstyle{^5}}}{r^{\textstyle{^6}}} \,\sin
 \left(2\left[\dot{\nu}-\dot{\theta}\right]\Delta t\right)
 +O(\inc^2/Q)+O(en/Q^2\chi)\;\;,~~
  \ea
 where the error $\,O(en/Q^2\chi)\,$ emerges when we identify the lagging
 distance $\,r(t-\Delta t)\,$ with $\,r\equiv r(t)\,$. Replacement of $\,r(t-\Delta t)\,$
 with $\,r\,$ is convenient, though not necessary. In subsection 7.2 below, we shall
 explain that, after averaging over one revolution of the moon about the planet, the
 error caused by this replacement reduces to $\,O(e^2n^2/Q^3\chi^2)\,$, which
 will be less than the largest error.

 The MacDonald torque (\ref{43}) is equivalent to the Darwin torque (\ref{32})
 with an important proviso that all time lags $\,\Delta
 t_{\textstyle{_{\it{l}mpq}}}\,$ are equal to one another or, equivalently,
 that the rheological model (\ref{model}) is accepted. Physically, the special
 case of equal time lags is exactly the case when the tide may be rigorously
 interpreted as one double bulge of a variable rate and amplitude.\footnote{~An
 attempt to generalise this simplified approach to arbitrary inclinations was
 undertaken by Efroimsky (2006). While for constant time lags that generalisation
 is likely to be acceptable, it remains to be explored whether it offers
 a practical approximation for actual rheologies (\ref{955}).}
 Mathematically, this model
 enables one to wrap up the infinite series (\ref{32}) into the elegant finite
 form (\ref{43}). Formally, this wrapping can be described like this: expression
 (\ref{43}) mimics the principal term of the series (\ref{32}), provided in this
 term the multiplier $\,G^2_{200}\,$ is replaced with unity, $\,a\,$ is replaced
 with $\,r\,$, and the principal phase lag
 \ba
 \epsilon_{\textstyle{_{\textstyle{_{2200}}}}}\,\equiv\,2\,(n\,-\,\dot{\theta}) \Delta t
 \label{}
 \ea
 is replaced with the longitudinal lag or, possibly better to say, with the quasi-phase
 \begin{equation}
 \epsilon\,\equiv\,2\,(\dot{\nu}\,-\,\dot{\theta})\,\Delta t\;\;\;.
 \label{11}
 \end{equation}
 Thus we see that within the MacDonald one-variable-bulge formalism the longitudinal lag
 (\ref{11}) is acting as an {\emph{instantaneous}} phase lag associated with the
 {\emph{instantaneous}} tidal frequency $\,\chi\,\equiv\,2\,|\dot{\nu}\,-\,\dot{\theta}|\,$.
 This is why we may call it simply $\,\epsilon\,$, without a subscript. Evidently,
 $\,\epsilon\,$ is (up to a sign) twice the geometrical angle subtended at the
 primary's centre between the directions to the moon and to the bulge.\footnote{~As the
 subtended angle is $\;|\,(\dot{\nu}\,-\,\dot{\theta})\,\Delta t\,|\;$, its double is equal
 to the absolute value of $\,\epsilon\,$, and not to that of
 $\,\epsilon_{\textstyle{_{\textstyle{_{220\mbox{\it{q}}}}}}}\,=\,2\,(n-\dot{\theta})\,\Delta t\,$.}

 The geometric meaning of the longitudinal lag being clear, let us consider its physical
 meaning, in the sense of this lag's relation to the dissipation rate. For some fixed
 frequency $\,\chi_{{\it l}mpq}\,$, the corresponding phase lag $\,\epsilon_{{\it l}mpq}\,
 $ is related to the appropriate quality factor via $\,1/Q_{{\it l}mpq}\,=\,\tan
 | \epsilon_{{\it l}mpq}|\,$. To keep the analogy between the true lags and the
 instantaneous lag (\ref{11}), one may conveniently {\emph{define}} a quantity $\,Q\,$ as
 the inverse of $\,\tan |\epsilon |\,$. This will enable
 one to express the MacDonald torque as
 \ba
 \nonumber
 \tau~=~\frac{3}{2}~GM_{sat}^2\;k_{2}
 \frac{R^{\textstyle{^5}}}{r(t)^{\textstyle{^{3}}}{r}(t-\Delta t)^{\textstyle{^{3}}}}\sin
 \epsilon+O(\inc^2/Q)
  ~~~~~~~~~~~~~~~~~~~~~~~~~~~~~~~~~~~\\
  \label{443}\\
  \nonumber
 =~\frac{3}{2}~{GM_{sat}^2}\;k_{2} \frac{R^{\textstyle{^5}}}{r^{\textstyle{^{6}}}
 }
 Q^{-1}\,\mbox{sgn}(\dot{\nu}-\dot{\theta})+O(\inc^2/Q)+O(en/Q^2\chi)+O(Q^{-3})
 ~~~.~~~~~~~~
 \ea
 Since $\,Q\,$ was {\emph{defined}} as $\,1/\tan |\epsilon |\,$, it is not
 guaranteed to deserve the name of an overall quality factor. At each
 particular frequency $\,\chi_{\textstyle{_{{\it l}mpq}}}\,$, the corresponding quality
 factor $\,Q_{{\it l}mpq}\,\equiv\,1/\tan |\epsilon_{{\it l}mpq}|\,$ is related to the peak
 energy of this mode,
 $\,E_{peak}(\chi_{\textstyle{_{{\it l}mpq}}})\,$, and to the one-cycle energy loss at
 this frequency, $\,\Delta E_{cycle}( \chi_{\textstyle{_{{\it l}mpq}}} )\,$, via
 \ba
 \Delta E_{cycle}(\chi_{\textstyle{_{{\it l}mpq}}})\;=\;-\;\frac{2\pi E_{peak}
 (\chi_{\textstyle{_{{\it l}mpq}}})}{
 Q_{\textstyle{_{{\it l}mpq}}}}\;\;\;.
 \label{820}
 \ea
 However, it is not at all obvious if the quantity $\,Q\,$ defined through the
 longitudinal lag as $\,Q\,\equiv\,1/\tan |\epsilon |\,$ interconnects the
 overall tidal energy with the overall one-cycle loss, in a manner similar to
 (\ref{820}). The literature hitherto has always taken for granted that it does.
 However, the proof (to be presented elsewhere) requires some effort. The proof
 is based on interpreting $\,\chi\,\equiv\,2\,|\dot{\nu}-\dot{\theta}|\,$ as an
 {\emph{instantaneous}} tidal frequency.

 The interconnection between $\,Q\,\equiv\,1/\tan |\epsilon |\,$ and the
 overall energy-damping rate mimics (\ref{820}) only up to a relative error
 of order $\,O(en/Q\chi)\,=\,O(en\,\Delta t)\,$, i.e., up to an absolute error of order
 $\,O(eQ^{-1}n\,\Delta t)\,$. This is acceptable, because in realistic settings $\;n\,\Delta t
 \,\ll\,1\;$.


 \subsection{Further simplifications available in the zeroth order of
 \it{\bf{\emph{en/Q{\chibold}}}}}

 Suppose we ignore the difference between $\,r\,$ and $\,r^*$, which are the two
 locations of the same satellite, separated by the time lag owing to the tidal
 response. We shall now demonstrate that, though the relative error of this
 approximations is $\,O(en/Q\chi)\,$, after averaging over a satellite period
 this approximation brings only a $\,O(e^2n^2/Q^2\chi^2)\,$ relative error into
 the expression for the torque.

 From the well-known formulae $\,r\,=\,a\,(1\,-\,e^2)/(1\,+\,e\,\cos\nu)\,$
 and $\,\partial\nu/\partial M\,=\,(1\,+\,e\,\cos\nu)^2/(1\,-\,e^2)^{3/2}\,$
 we see that
 \ba
 \nonumber
 \Delta r\equiv r(t)-r(t-\Delta t)=-\frac{a\,e\,(1\,-\,e^2)}{(1\,+\,e\;\cos\nu)^2}\;
 \sin\nu\;\Delta\nu\,+\,O\left(e\,(\Delta\nu)^2\,\right)\,=\,
 -\,\frac{a\,e\;\sin \nu}{(1-e^2)^{1/2}}\,n\,\Delta t\,+\,O\left(e\,(n\;\Delta t)^2\,\right)~~~.
 \ea
 The time lag is interconnected with the phase shift and the
 quality factor via the relations
 \ba
 \nonumber
 \chi\;\Delta t\;=\;\epsilon\;\approx\;Q^{-1}\;\;\;,
 \label{}
 \ea
 $\,\chi\,=\,2\,|\dot{\theta}\,-\,\dot{\nu}\,|\,$ being the instantaneous tidal frequency.
 Hence
 \ba
 \nonumber
 \Delta r\,\equiv\,r(t)\,-\,r(t-\Delta t)\,\approx\;-\;a\;\frac{e}{Q}\;\frac{n}{\chi}\;
 \sin\nu
 ~~~.
 \ea
 %
 As $\,\Delta r\,$ is proportional to $\,\sin\nu$, only terms quadratic in
 $\,\Delta r\,$ survive averaging. Thus, while in
 \ba
 U\,=\,-\,\frac{3}{4}\,{G\,M_{sat}^*}\,k_{2}\,
 \frac{R^{\textstyle{^5}}}{r^6} \,\cos (2\lambda-2\lambda^* +
 \epsilon)+O(en/Q\chi)+O(\inc^2)+O({\inc^*}^2)+O(\inc\inc^*)~~,~~~
 \label{44}
 \ea
 and
 \ba
 \tau~=~\frac{3}{2}\,{G\,M_{sat}\,M_{sat}^*}\,
 k_{2}\,\frac{R^{\textstyle{^5}}}{
 r^6} \,\sin (2\lambda-2\lambda^* +
 \epsilon)\;
 +\,O(en/Q\chi)\,
 +\,O(\inc^2)\,~~,~~~
 \label{45}
 \ea
 the {\emph{relative}} error is $\,O(en/Q\chi)+O(\inc^2)\,$, in the averaged
 expression\footnote{~We recall that time averages over one revolution of the
 satellite about the primary are given by
 \ba
 \nonumber
 \langle\;\,.\,.\,.\,\;\rangle\;\equiv\;
 \frac{1}{2\;\pi}\;\int_{0}^{2\pi}\;.\,.\,.\;\;\;d{\cal{M}}
 \;=\;\frac{\left(1\;-\;e^2\right)^{3/2}}{2\;\pi}\;
 \int_{0}^{2\pi}\;.\,.\,.\;\;\;\frac{d\nu}{\left(1\;+\;e\;\cos \nu  \right)^2}
 \;\;\;\;,
 \ea
 while the planetocentric distance is
 $\,r=a\left(1-e^2\right)/\left(1+e\,\cos \nu\right)\,$,
 with $\nu$ being the true anomaly. This way,
  \ba
 \nonumber
 \langle\;\,\frac{R^{\textstyle{^6}}}{r^6}\;\,
 \sin\epsilon
 \,\;\rangle\;=\;
 \frac{\left(1\;-\;e^2\right)^{3/2}}{2\;\pi}\;\int_{0}^{2\pi}\;
 \frac{R^{\textstyle{^6}}}{r^6}\;\;
 \sin\epsilon
 \,\;\frac{d\nu}{\left(1\;+\;e\;
 \cos \nu  \right)^2}~~~~~~~~~~~~~~~~~~~~~~~~~~~~~~~~~~~~~\\
 \nonumber\\
 \nonumber\\
 \nonumber
 =\;\frac{\left(1\;-\;e^2\right)^{3/2}}{2\;\pi}\;\int_{0}^{2\pi}\;
 \frac{R^{\textstyle{^6}}}{r^6}\;\;
 \sin\epsilon
 \,\;\frac{r^2\;\;d\nu}{a^2\;
 \left(1\;-\;e^2\right)^{2}}\;
 =\;\frac{R^{\textstyle{^2}}}{a^2}\;\frac{1}{2\,\pi\,\;\left(1\;-\;e^2
 \right)^{1/2}\,}\;\int_{0}^{2\pi}\;\frac{R^{\textstyle{^4}}}{r^4}\;\;
 \sin\epsilon
 \,\;{d\nu}\;\;\;\;.
 \ea
 }
 \begin{subequations}
 \label{446}
 \ba
 \label{446a}
 \langle\,\tau
 \,\rangle~=
 ~-~\frac{3\,G\,M_{sat}^{\textstyle{^{\,2}}}\,k_{2}}{2\;R}\;\;\langle\;\,
 \frac{R^{\textstyle{^6}}}{r^6}\;\,
 \sin\epsilon
 \,\;\rangle~\,+\,O(e^2n^2/Q^3\chi^2)\,
 +\,O(\inc^2/Q)~~~~~~~~~~~~~~~~~~~~~~~\\
 \nonumber\\
 \nonumber\\
 \label{446b}
 =\;-~\frac{3\,G\,M_{sat}^{\textstyle{^{\,2}}}\,k_{2}\,R}{4\;\pi\;a^2}
 \;\,\frac{1}{\,\left(1\;-\;e^2\right)^{1/2}\,}\;\int_{0}^{2\pi}\;
 \frac{R^{\textstyle{^4}}}{r^4}\;\,
 \sin\epsilon\;\;{d\nu}\,\,+\,O(e^2n^2/Q^3\chi^2)\,
 +\,O(\inc^2/Q)\;~~.~~~~~~~~
 \ea
 \end{subequations}
 it is only $\,O(e^2 n^2/Q^2\chi^2)+O(\inc^2)\,$.

 In the above expressions, we asserted after the differentiation that $\,M_{sat}^*\,
 =\,M_{sat}\,$ and $\,\lambda^*\,=\,\lambda\,$, implying that the tide-generating and
 tidally-perturbed moons are one and the same body. As soon as $\,\lambda\,$ is
 set to be equal to $\,\lambda^*\,$, the {{sine}} function in (\ref{446})
 becomes $\,\sin\epsilon\,\approx\,1/Q\,$. So, while the {\emph{relative}}
 error in (\ref{446}) is $\;O(e^2n^2/Q^2\chi^2)\,+\,O(\inc^2)\;$, the
 {\emph{absolute}} error becomes $\;O(e^2n^2/Q^3\chi^2)\,+\,O(\inc^2/Q)\;$.


 The error $\;O(e^2n^2/Q^3\chi^2)\;$ becomes irrelevant for two reasons. First,
 our substitution of $\,\sin\epsilon\,$ with $\,\tan\epsilon\,=\,1/Q\,$ generates
 a relative error $\,O(Q^{-2})\,$, i.e., an absolute error $\,O(Q^{-3})\,$.
 Second, as explained in the end of subsection 7.1, the uncertainties inherent in our
 definition of the overall quality factor $\,Q\,$ entail an absolute error
 $\;O(en/Q^2\chi)\;$. Each of
 these two errors exceeds $\;O(e^2n^2/Q^3\chi^2)\;$. We can then write:
 \begin{subequations}
 \label{46}
 \ba
 \label{46a}
 \langle\,\tau
 \,\rangle~=
 ~-~\frac{3\,G\,M_{sat}^{\textstyle{^{\,2}}}\,k_{2}}{2\;R}\;\;
 \langle\;\;\frac{\mbox{sgn}(\dot{\theta}\,-\;\dot{\nu})}{
 Q}\,\;
 \frac{R^{\textstyle{^6}}}{r^6}\;\;\rangle~\,+\,O(Q^{-3})\,
 +\,O(\inc^2/Q)\,+\,O(en/Q^2\chi)~~~~~~~~~~~~~~~~~\\
 \nonumber\\
 \nonumber\\
 \label{46b}
 =\,-\,\frac{3\,G\,M_{sat}^{\textstyle{^{\,2}}}\,k_{2}\,R}{4\;\pi\;a^2}
 \;\,\frac{1}{\,\left(1\;-\;e^2\right)^{1/2}\,}\;\int_{0}^{2\pi}\;
 \frac{R^{\textstyle{^4}}}{r^4}\;\,\frac{\mbox{sgn}(\dot{\theta}\,-\;\dot{\nu})}{
 Q}
 \,\;{d\nu}\,\,+\,O(Q^{-3})\,+\,O(\inc^2/Q)\,+\,O(en/Q^2\chi)\;~.~~~~
 \ea
 \end{subequations}

 \section{Use and abuse of approximation (\ref{44} - \ref{46})}

 Just as with the formula (\ref{44}) for the potential, the elegant expression (\ref{45})
 for the torque remain correct only to the zeroth order in $\,e/Q\,$, while (\ref{46})
 is valid to the first order. This is the reason why the convenience of this
 approximation and of its corollaria is somewhat deceptive. Nevertheless, the
 (\ref{44}) - (\ref{46}) were employed by many an author.

 Goldreich \& Peale (1966) used them to build a theory containing
 terms up to $\,e^7\,$. Now we see that some coefficients in their theory
 of capture into resonances must be reconsidered. The same pertains to
 some coefficients in the theory of Mercury's rotation, recently offered
 by Peale (2005). Fortunately, the key conclusions of Peale (2005) stay unaltered,
 despite the corrections needed in the said coefficients.\footnote{~Stan Peale,
 private communication, 2007.}

 Interestingly, Kaula (1968) fell into this temptation, and so did Goldreich
 (1966b). Equation (4.5.29) in Kaula (1968), as well as equation (15) in
 Goldreich (1966b), is but the above formula (\ref{46}) with the inverse
 quality factor taken out of the integral:
 \ba
 \nonumber
 \tau^{Kaula}\;=\;-~\frac{3\,G\,M_{sat}^{\textstyle{^{\,2}}}\,k_{2}\,R}{4\;\pi\;a^2\;Q}
 \;\,\frac{1}{\,\left(1\;-\;e^2\right)^{1/2}\,}\;\int_{0}^{2\pi}\;
 \frac{R^{\textstyle{^4}}}{r^4}\;\,{\mbox{sgn}(\dot{\theta}\,-\,\,\dot{\nu})}
 \,\;{d\nu}\;~~.~~~~~~~~~~(\mbox{Kaula 1968, eqn$\,$4.5.29})
 \ea
 Besides the afore-mentioned fact that this approach contains a relative error
 $\,O(en/Q\chi)\,+\,O(Q^{-2})+O(\inc^2)\,$, it suffers a greater defect. Taking
 $Q^{\textstyle{^{-1}}}$ out of the integral is illegitimate, because it
 implies frequency-independence of $\,Q\,$. This is then incompatible with
 Kaula's implicit assumption of a constant $\,\Delta t\,$, an assumption
 tacitly present in (\ref{trick}).\footnote{~This would be incompatible with the
 MacDonald (1964) treatment as well, because MacDonald's formalism
 necessitates the rheology $\,Q\sim1/\chi\,$. We
 shall return to this point in subsection 10.1.}

 Goldreich (1966b) and Kaula (1968) used this oversimplified formula to
 investigate librations of a satellite trapped in a 1:1 resonance.
 Other authors used it to evaluate despinning rates of bodies outside this
 resonance. We shall dwell on the latter case in section 10.

 \section{Can the quality factor scale as a positive power of
the tidal frequency?}

 As of now, the functional form of the dependence $\,Q(\chi)\,$ for Jovian
 planets remains unknown. For terrestrial planets, the model $\,Q\,\sim\,1/\chi\,
 $ is definitely incompatible with the geophysical data. A convincing volume of
 measurements firmly witnesses that $\,Q\,$ of the mantle scales as the tidal
 frequency to a {\emph{positive}} fractional power:
 \ba
 Q
 \;=\;\left(\,{\cal E}\,\chi\,\right)^{\textstyle{^{\alpha}}}
 \;\;\;,
 ~~~~~\mbox{where}\;\;\;\alpha\,=\,0.3\,\pm\,0.1\;\;\;,
 \label{955}
 \ea
 ${\cal E}\,$ being an integral rheological parameter with dimensions of time.
 This rheology is incompatible with the postulate of frequency-independent
 time-delay. Therefore, insertion of the realistic model (\ref{955}) into
 the formula presented in section 8 will remain insufficient. An honest
 calculation should be based on averaging the Darwin-Kaula-Goldreich formula
 (\ref{exp}), with the actual scaling law (\ref{955}) inserted therein, and
 with the appropriate dependence $\,\Delta t_{\textstyle{_{lmpq}}}(\,\chi_{\textstyle{_{lmpq}}}\,)\,$
 taken into account (see formula (\ref{631}) below).

  \subsection{The ``paradox"}

 Although among geophysicists the scaling law (\ref{955}) has long become common knowledge,
 in the astronomical community it is often met with prejudice. The prejudice stems from
 the fact that, in the expression for the torque, $\,Q\,$ stands in the denominator:
 \ba
 \tau\;\sim\;\frac{1}{Q}\;\;\;\,.\,~~~~~~~~~~~~~~~~~
 \label{torque}
 \ea
 At the instant of crossing the synchronous orbit, the principal tidal frequency
 $\,\chi_{\textstyle{_{\textstyle{_{2200}}}}}\,$ becomes nil, for which reason
 insertion of
 \ba
 Q
 \;\sim\;\chi^{\alpha}
 \;\;\;,
 ~~~~~\alpha\,>\,0\;\;\;
 \label{}
 \ea
 into (\ref{torque}) seems to entail an infinitely large torque at the
 instant of crossing:
 \ba
 \tau\;\sim\;\frac{1}{Q}\;\sim\;\frac{1}{\chi^{\alpha}}\;\rightarrow\;
 \infty
 ~~,~~~~~{\mbox{for}}~~~\chi\;\rightarrow\;0~~~,
  \label{infinity}
 \ea
 a clearly unphysical result.

 Another, very similar objection to (\ref{955}) originates from the fact that
 the quality factor is inversely proportional to the phase shift: $\,Q\,\sim\,
 1/\epsilon\,$. As the shift (\ref{24}) vanishes on crossing the synchronous orbit,
 one may think that the value of the quality factor must, effectively, approach
 infinity. On the other hand, the principal tidal frequency vanishes on crossing
 the synchronous orbit, for which reason (\ref{955}) makes the quality factor
 vanish. Thus we come to a contradiction.

 For these reasons, the long-entrenched opinion is that these models
 introduce discontinuities into the equations and can thus be considered as
 unrealistic approximations for rotating bodies.

 It is indeed true that, while law (\ref{955}) works over scales shorter than
 the Maxwell time (about $\,10^2$ yr for most minerals), it remains subject
 to discussion in regard to longer timescales. Nonetheless, it should be
 clearly emphasised that the infinities emerging at the synchronous-orbit
 crossing can in no way disprove any kind of rheological model. They can
 only disprove the
 flawed mathematics whence they provene.

 \subsection{A case for reasonable doubt}

 To evaluate the physical merit of the alleged infinite-torque ``paradox",
 recall the definition of the quality factor. As part and parcel of the
 linearity approximation, the overall damping inside a body is expanded
 in a sum of attenuation rates corresponding to each periodic disturbance:
 \ba
 \langle\,\dot{E}\;\rangle\;=\;\sum_{i}\;\langle\,
 \dot{E}(\chi_{\textstyle{_i}})\;\rangle
 \label{407}
 \ea
 where, at each frequency $\,\chi_i\,$,
 \ba
 \langle\,\dot{E}(\chi_{\textstyle{_i}})~\rangle~=~-~2~\chi_{\textstyle{_i}}~
 \frac{\,\langle\,E(\chi_{\textstyle{_i}})~\rangle\,}{Q(\chi_{\textstyle{_i}})
 }~=\,\;-\;\chi_{\textstyle{_i}}\;\frac{\,E_{_{peak}}(\chi_{\textstyle{_i}})
 \,}{Q(\chi_{\textstyle{_i}})}\;\;\;,
 \label{408}
 \ea
 $\langle\,.\,.\,.\,\rangle~$ designating an average over a flexure cycle,
 $\,E(\chi_{\textstyle{_i}})\,$ denoting the energy of deformation at
 the frequency $\,\chi_{\textstyle{_i}}\,$, and $Q(\chi_{\textstyle{_i}})\,$
 being the quality factor of the medium at this frequency.

 This definition by itself leaves enough room for doubt in the above ``paradox".
 As can be seen from (\ref{408}), the dissipation rate is proportional not to
 $\;1/Q(\chi)\;$ but to $\;\chi/Q(\chi)\;$. This way, for the dependence $\,Q
 \,\sim\,\chi^{\alpha}\,$, the dissipation rate $\,\langle\dot{E}\rangle\,$
 will behave as $\,\chi^{1-\alpha}\,\;$. In the limit of $\,\chi\,\rightarrow
 \,0\,$, this scaling law portends no visible difficulties, at least for the
 values of $\,\alpha\,$ up to unity. While raising $\,\alpha\,$ above unity may
 indeed be problematic, there seem to be no fundamental obstacle to having
 materials with positive $\,\alpha\,$ taking values up to unity. So far, such
 values of $\,\alpha\,$ have caused no paradoxes, and there seems to be no reason
 for any infinities to show up.


 \subsection{The phase shift and the quality factor}

 As another preparatory step, we recall that, rigorously speaking, the
 torque is proportional not to the phase shift $\,\epsilon\,$ itself but
 to $\,\sin\epsilon\,$. From (\ref{34}) and (\ref{955}) we obtain:
 \ba
 |\,\sin\epsilon\,|\;=\;\frac{
 1}{\textstyle\sqrt{1\,+\,Q^{\textstyle{^2}}}}\;=\;\frac{1}{\sqrt{1\,+\;{\cal E}^{
 \textstyle{^{2
 \alpha}}}\;{\chi}^{\textstyle{^{2\alpha}}}\;}}\;\;\;.
 \label{shift}
 \ea
 We see that only for large values of $\,Q\,$ one can approximate $\;\,|\,
 \sin\epsilon\,|\;\,$ with $\,1/Q\,$ (crossing of the synchronous orbit
 {\emph{not}} being the case). Generally, in any expression for the
 torque, the factor $\,1/Q\,$ must always be replaced with $\,1/\sqrt{1\,+\,
 Q^2}\,\;$. Thus instead of (\ref{torque}) we must write:
 \ba
 \tau\;\sim\;|\,\sin\epsilon\,| \;=\;\frac{1}{\sqrt{1\,+\,Q^{\textstyle{^2}}\;}}\;=\;\frac{1}{\sqrt{1\,+\;{\cal E}^{
 \textstyle{^{2
 \alpha}}}\;{\chi}^{\textstyle{^{2\alpha}}}\;}}\;\;\;\;\,,\,~~~~~~~~~~~~~~~~~
 \label{relation}
 \ea
 ${\cal E}\,$ being a dimensional constant from (\ref{955}).

 Although this immediately spares us from the fake infinities at $\chi
 \rightarrow 0$, we still are facing a strange situation: it follows
 from (\ref{shift}) that, for a positive $\alpha$ and vanishing
 $\chi$, the phase lag $\epsilon$ must be approaching $\pi/2$,
 thereby inflating the torque to its maximal value (while on physical
 grounds the torque should vanish for zero $\chi$). Evidently, some
 important details are still missing from the picture.

  \subsection{The stone rejected by the builders}


 To find the missing link, recall that Kaula (1964) described tidal damping
 by employing the method suggested by Darwin (1880): he accounted for
 attenuation by merely adding a phase shift to every harmonic involved --
 an empirical approach intended to make up for the lack of a consistent
 hydrodynamical treatment with viscosity included. It should be said,
 however, that prior to the work of 1880 Darwin had published a less known
 article (Darwin 1879), in which he attempted to construct a
 self-consistent theory, one based on the viscosity factor of the mantle,
 and not on empirical phase shifts inserted by hand. Darwin's conclusions
 of 1879 were summarised and explained in a more general mathematical
 setting by Alexander (1973).

 The pivotal result of the self-consistent hydrodynamical study is the
 following. When a variation of the potential of a tidally distorted planet,
 $\,U(\erbold)\,$, is expanded over the Legendre functions $\,P_{{\it{l}}m}(
 \sin\phi)\,$, each term of this expansion will acquire not only a phase lag
 but also a factor describing a change in amplitude. This forgotten factor,
 derived by Darwin (1879), is nothing else but $\;\,\cos\epsilon\;$. Its
 emergence should in no way be surprising if we recall that the damped,
 forced harmonic oscillator
 \ba
 \ddot{x}\;+\;2\;\gamma\;\dot{x}\;+\;\omega^2_o\,x\;=\;F\;e^{\inc\,\lambda\,t}
 \label{}
 \ea
 evolves as
 \ba
 x(t)~=~C_1~\,e^{\textstyle{^{(\,-\,\gamma\,+\,\inc\,\sqrt{\omega_o^2-\gamma^2
 \,}\,)\;t}}}\,+\;
 C_2~\,e^{\textstyle{^{(\,-\,\gamma\,-\,\inc\,\sqrt{\omega_o^2-\gamma^2\,}\,)
 ~t}}}\,+~\frac{F~\cos\epsilon}{\omega_o^2\,-\,\lambda^2}~\,e^{\textstyle{^{
 \inc\,(\lambda\,t\,-\,\epsilon)}}}\;\;\;,
 \label{solution}
 \ea
 where the phase lag is
 \ba
 \tan\epsilon\;=\;\frac{2\;\gamma\;\lambda}{\left(\,\omega_o^2\,-\;\lambda^2\,\right)}
 \;\;\;,
 \label{}
 \ea
 and the first two terms in (\ref{solution}) are damped away in
 time.\footnote{~As demonstrated by Alexander (1973), this example indeed has relevance to
 the hydrodynamical theory of Darwin, and is not a mere illustration. Alexander (1973)
 also explained that the emergence of the $\,\cos\epsilon\,$ factor is generic. (Darwin
 (1879) had obtained it in the simple case of $\,{\it l}\,=\,2\,$ and for a special value
 of the Love number: $\,k{\it{_{2}}}=\,1.5\,$.)

 A further investigation of this issue was undertaken in a comprehensive work
 by Churkin (1998), which unfortunately has never been published in English
 because of a tragic death of its Author. In this preprint, Churkin explored the
 frequency-dependence of both the Love number $\,k_2\,$ and the quality factor
 within a broad variety of rheological models, including those of Maxwell and
 Voight. It follows from Churkin's formulae that within the Voigt model the
 dynamical $\,k_2\,$ relates to the static one as $\,\cos \epsilon\,$. In the
 Maxwell and other models, the ratio approaches $\,\cos \epsilon\,$ in the
 low-frequency limit.}

 In the works by Darwin's successors, the allegedly irrelevant factor of $\,\cos
 \epsilon\,$ fell through the cracks, because the lag was always asserted to be
 small. In reality, though, each term in the Fourier expansions (\ref{21}),
 (\ref{30} - \ref{33}), and (\ref{35}) should be amended with $\,\cos\epsilon_{
 \textstyle{_{\textstyle{_{{\it{l}}mpq}}}}}\,$. Likewise, the correct versions of
 (\ref{42} - \ref{43}) and (\ref{443}) should contain an extra factor of $\,\cos\epsilon_{
 \textstyle{_{\textstyle{_{2200}}}}}\,$. For the same reason, instead of
 (\ref{relation}), we should write down:
 \ba
 \tau\;\sim\;|\,\cos\epsilon\;\,\sin\epsilon\,|\;=\;\frac{Q}{\sqrt{1\,+\,
 Q^{\textstyle{^2}}\;}}\;\frac{1}{\sqrt{1\,+\,Q^{\textstyle{^2}}\;}}\;=\;\frac{{\cal E}^{\textstyle{^{
 \alpha}}}\;{\chi}^{\textstyle{^{\alpha}}}}{1\,+\;{\cal E}^{\textstyle{^{2
 \alpha}}}\;{\chi}^{\textstyle{^{2\alpha}}}}\;\;\;\;\,.\,~
 \label{}
 \ea
 At this point, it would be tempting to conclude that, since (71) vanishes in the limit of
 $\chi\rightarrow0\;$, {\emph{for any sign}} {\emph{of}} $\alpha\;$, then no paradoxes happen on the satellite's crossing the
 synchronous orbit. Sadly, this straightforward logic would be too
simplistic.

 In fact, prior to saying that $\,\cos\epsilon\,\sin\epsilon\rightarrow0$, we must take into consideration one more
 subtlety missed so far. As demonstrated in the Appendix, taking the limit of $Q\rightarrow 0$ is a
 nontrivial procedure, because at small values of $\,Q\,$ the interconnection between the lag
 and the Q factor becomes very different from the conventional $Q=\cot|\epsilon|$. A laborious
 calculation shows that, for $\;Q\,<\,1-\pi/4\,$, the relation becomes:
  \ba
  \nonumber
  |\,\sin\epsilon\,\cos\epsilon\,|\,=\;(3Q)^{1/3}\,\left[1-\frac{4}{5}(3Q)^{2/3}+O(Q^{4/3})\right]\;\;\;,
  \label{}
  \ea
 which indeed vanishes for $Q\rightarrow 0$. Both $\,\epsilon_{\textstyle{_{2200}}}\,$ and the
 appropriate component of the torque change their sign on the satellite crossing the synchronous orbit.

 So the main conclusion remains in force: nothing wrong happens on crossing the synchronous orbit, ~Q.E.D.

   \section{Tidal despinning.}

 The following formula for the average deceleration rate $\,\ddot{\theta}\,$ of
 a planet
 due to a tide-raising satellite has often appeared in the literature:
 \ba
 \langle\;\ddot{\theta}\;\,\rangle\;=\;-\;{\cal K}\;\left[
 \;\dot{\theta}\;\,{\cal A}(e)\;-\;n\;{\cal N}(e)\;\right]\;\;\;,
 \label{Laskar}
 \label{647}
 \label{621}
 \ea
 where
 \ba
 {\cal A}(e)\;=\;\left(\,1\;+\;3\;e^2\;+\;\frac{3}{8}\;e^4\,
 \right)\;\left(\,1\,-\,e^2\,\right)^{-9/2}~~~,~~~~~~~~~~~~~
 \label{648}
 \label{622}
 \ea
 and
 \ba
 {\cal N}(e)\;=\;\left(\,1\;+\;\frac{15}{2}\;e^2\;+\;\frac{45}{8}\;
 e^4\;+\;\frac{5}{16}\;e^6\,\right)\;\left(\,1\,-\,e^2\,\right)^{-6}~~~,
 \label{649}
 \label{623}
 \ea
 $\theta\,$ being the sidereal angle, $\,\dot{\theta}\,$ being the primary's
 spin rate, $\,{\cal K}\,$ being some constant, and the angular brackets
 designating an average over one revolution of the secondary about the primary.
 This expression was derived by different methods in Goldreich \& Peale (1966)
 and Hut (1981), and was later employed by Dobrovolskis (1995, 2007) and
 Correia \& Laskar (2004)\footnote{~Our formula (78) differs from formula (4) in Correia and Laskar
 (2004) by a factor of $\,n/{\chi}\,$, because in {\it{Ibid.}} the quality factor had been introduced
 as $\, 1/( n\, \Delta t )\,$ and not as $ \,1/( \chi\, \Delta t )\, $.}.

 \subsection{{Derivation by means of the MacDonald torque}}

 The following proof of (\ref{647} - \ref{649}) is implied in Goldreich \&
 Peale (1966) and is presented in more detail in Dobrovolskis (2007). Their
 starting point was the MacDonald torque (\ref{46}). Hut (1981), who approached
 the issue in the language of the Lagrange-type planetary equations, took into
 account, in the disturbing function, only the leading term of series
 (\ref{21}), and substituted the principal tidal frequency
 $\,\chi_{\textstyle{_{2200}}}\,=\,2\,|\dot{\theta}\,-\,n|\,$ with the synodal frequency
 $\,\chi\,=\,2\,|\dot{\theta}\,-\,\dot{\nu}|\,$. Thereby, his approach was
 equivalent to that of Goldreich \& Peale (1966) and Dobrovolskis (2007).

 Although not necessarily assumed by these authors,\footnote{~It should be
 mentioned that the original treatment by MacDonald (1964) is inherently
 contradictory.  On the one hand, MacDonald postulates
 that there exists one overall double bulge. As explained in subsection
 7.1 above, this assertion unavoidably implies constancy of the time lag
 $\,\Delta t\,$, so that $\,Q\sim1/\chi\,$ and $\,\epsilon\sim\chi\,$.
 However, MacDonald (1964) erroneously set $\,Q\,$ (and, thence, also
 $\,\epsilon\,$) frequency-independent, an assertion incompatible with his
 postulate of existence of an overall double bulge.

 Whenever in the current paper we refer to MacDonald's torque, we always imply
 his postulate that one double bulge exists. At the same time, to make
 the MacDonald
 treatment consistent, we always adjust the
 MacDonald
 treatment by letting $\,Q\,$ and $\,\epsilon\,$ scale as
 $\,1/\chi\,$ and $\,\chi\,$, correspondingly.} their method, as we saw in the
 section above, inherently implied the following assertions:\\

 (I) ~~~The quantity $\,\chi\,=\,2\,|\dot{\theta}\,-\,\dot{\nu}|\,$ is treated
 as an instantaneous tidal frequency. Accordingly, the overall quality factor
 $\,Q\,$ is implied to be a function not of the principal frequency $\,
 \chi_{\textstyle{_{2200}}}\,$ but of the instantaneous frequency $\,\chi\,$.\\

 (II) ~~The functional form of this dependence is chosen as
 $\;Q\,=\,{(\Delta t)^{-1}\;\chi^{-1}}\;$, where $\Delta t\,$ is the time lag.\\

 (III) ~The time lag $\,\Delta t\,$ is frequency-independent. This assertion is
 equivalent to (II), as can be demonstrated from (\ref{24}).\\

 Beside this, those authors neglected the order-$en/Q\chi\,$ difference between
 $\,r\,$ and $\,r^*$ in (\ref{443}), generating a relative error in $\,\tau\,$
 of order $\,O(en/Q\chi)\,$
 (which, luckily, reduced to $\,O(e^2n^2/Q^2\chi^2)\,$ after orbital averaging).
 They also substituted $\,\sin \epsilon\,$ with $\,1/Q\,$, causing a relative
 error of order $\,O(1/Q^2)\,$, because in reality $\,Q\,$ is the reciprocal of
 $\,\tan\epsilon\,$, not of $\,\sin\epsilon\,$.\\

 Assertion (II) can be written down in more generic notation:
 \ba
 Q\;=\;\left(\,{\cal E}\,\chi\,\right)^{\textstyle{^{\alpha}}}~~~,~~~\mbox{with}~~~\alpha\;=\;-\;1~~~.
 \label{650}
 \label{624}
 \ea
 This form of the scaling law is more convenient, for it leaves one an opportunity
 to switch to different values of $\,\alpha\,$. For any value of $\,\alpha\,$ (not
 only for $\,-1\,$), the constant ${\cal E}\,$ is an integral rheological parameter (with
 the dimension of time), whose physical meaning is explained in Efroimsky \& Lainey
 (2007). It can be shown that in the particular case of $\;\alpha\,=\,-\,1\;$ the
 parameter $\,{\cal E}\,$ coincides with $\,\Delta t\,$. In realistic situations,
 $\,\alpha\,$ differs from $\,-1\,$, while the parameter $\,{\cal E}\,$ is related to the
 time lag in a more sophisticated way ({\emph{Ibid.}}).

 To show how (\ref{Laskar} - \ref{649}) stem from the above Assertions, keep
 for the time being $\,\alpha=\,-\,1\,$. Also recall that the torque is
 despinning (so $\,\dot{\theta}\,>\,n\,$), and that for the averages over time
 \ba
 \langle\;\ddot{\theta}\;\,\rangle\;=\;
 \frac{\langle\,\tau\,\rangle}{C}\;\;\;,
 \label{}
 \ea
 $C\,=\,\xi\,M_{planet}\,R^2\,$ being the maximal moment of inertia of the planet. (For
 a homogeneous spherical planet, $\xi=2/5$.) Then plug (\ref{650}) into (\ref{46}) and
 average the torque:\footnote{~As explained in the paragraph preceding formula (\ref{46}),
 substitution of $\,\sin\epsilon\,$ with $\,1/Q\,$ in the expression for torque generates
 a relative error $\,O(Q^{-2})\,$, i.e., an absolute error $\,O(Q^{-3})\,$. Instead of
 inserting (\ref{650}) into (\ref{46}), one may directly use (\ref{11}). Still,
 approximation of $\,\sin\epsilon\,$ with $\,\epsilon\,$ will entail, in (\ref{653}) and
 its corollaria, a relative error $\,O(Q^{-2})\,$ and an absolute error
 $\,O(Q^{-3})\,$. The situation will become more complicated in the special case of
low values of the quality factor. See the Appendix below}
 \begin{subequations}
 \label{653}
 \ba
 \nonumber
 \langle\,\tau\,\rangle=
 \,-\;\frac{3\,G\,M_{sat}^{\textstyle{^{\,2}}}\;\,k_{2}\;{\cal E}}{R}\;\;
 \langle\;\,(\dot{\theta}\,-\,\dot{\nu})\,\;
 \frac{R^{\textstyle{^6}}}{r^6}\;\;\rangle~~+O(\inc^2/Q)+O(Q^{-3})+O(en/Q^2\chi)
 \;=~~~~~~~~~~~~~~~~~~~~~~~~~~
 \label{}
 \ea
 \ba
 -\;\frac{3\,G\,M_{sat}^{\textstyle{^{\,2}}}\;\,k_{2}\;{\cal
 E}}{R}\;\dot{\theta}\;\;\langle\,
 \frac{R^{\textstyle{^6}}}{r^6}\;\rangle\;+\;\frac{3\,G\,M_{sat}^{\textstyle{^{\,2}}}\;
 k_{2}\;{\cal E}}{R}\;\langle\;\dot{\nu}\,\frac{R^{\textstyle{^6}}}{r^6}\;\rangle+
 O(\inc^2/Q)+O(Q^{-3})+O(en/Q^2\chi)~~~~~
 \label{653a}
 \ea
 \ba
 \nonumber
  =&-&\frac{3\,G\,M_{sat}^{\textstyle{^{\,2}}}\;\,k_{2}\;{\cal E}}{R}\;\dot{\theta}\;\;
 \frac{R^6}{a^6}\left(1\,-\,e^2\right)^{-9/2}~\,\frac{1}{2\,\pi}\;
 \int_{0}^{2\pi}\left(1+e\;\cos\nu\right)^4\,d\nu\;~~~~
 ~~~~~~~~~~~~~~~~~~~~~\\
 \nonumber\\
 \nonumber\\
 &+&\frac{3\,G\,M_{sat}^{\textstyle{^{\,2}}}\;k_{2}\,{\cal E}}{R}\,n\,\frac{R^{\textstyle{^6}}}{a^6}
 \left(1-e^2\right)^{-6}\frac{1}{2\pi}\int_{0}^{2\pi}\left(1+e\,\cos\nu\right)^6
 d{\nu}+O(\inc^2/Q)+O(Q^{-3})+O(en/Q^2\chi)~~,~~~~~~~~~~~~~~
 \label{653b}
 \ea
 \end{subequations}
 where the absolute error $\,O(en/Q^2\chi)\,$ emerges  due to an uncertainty in the definition of
 the overall quality factor $\,Q\,$ employed in MacDonald's model.

 Evaluation of the above integrals is trivial and indeed leads to (\ref{647} -
 \ref{649}), the constant being
 \ba
 {\cal K}\,=\,\frac{3\,G\,M_{sat}^{\textstyle{^{\,2}}}\;\,k_{2}\;{\cal E}}{C\;R}\;
 \frac{R^{\textstyle{^6}}}{a^6}\,
 =\,\frac{3\,n^2\,M_{sat}^{\textstyle{^{\,2}}}\;\,k_{2}\;{\Delta t}}{\xi\;
 M_{planet}\;(M_{planet}\,+\,M_{sat})}\;\frac{R^{\textstyle{^3}}}{a^3}
 \,=\,\frac{3\,n\,M_{sat}^{\textstyle{^{\,2}}}\;\,k_{2}}{\xi\;Q\;M_{planet}\;(M_{planet}
 \,+\,M_{sat})}\;\frac{R^{\textstyle{^3}}}{a^3}\;\,
 \frac{n}{\chi}\;\;\;,~~~
 \label{654}
 \ea
 where we used the fact that for $\,\alpha=-1\,$ the rheological parameter
 ${\cal E}$ is simply the lag $\,\Delta t$.


 It should also be added that, since (\ref{653b}) contains a relative
 error $\,O(Q^{-2})\,$, the usefulness of the $\,e^4\,$ and $\,e^6\,$ terms in
 (\ref{648} - \ref{649}) depends on the values of the eccentricity and the
 quality factor. If, for example, $\,Q=70\,$, then the $\,e^4\,$ terms become
 unimportant for $\,e<0.12\,$, while the $\,e^6\,$ terms become unimportant for
 $\,e<0.24\,$.

 To draw to a close, we would mention that besides the above
 formula (\ref{Laskar}), in the literature hitherto we saw its sibling,
 an expression derived in a similar way, but with Assertion II rejected in favour of
 treating $\,Q\,$ as a frequency-independent constant. The result of this treatment
 suffers an incurable birth trauma -- the incompatibility between the
 frequency-independence of $\,\Delta t\,$ and the frequency-independence of $\,Q\,$.

 \subsection{Calculation based on the Darwin torque}

 The following alternative derivation is based on the same Assertions (I - III) and,
 naturally, leads to the same results. The idea is to calculate the despinning rate not in
 terms of the MacDonald torque, but in terms of the Darwin torque, keeping the
 eccentricity-caused relative error at the level of $\,O(e^6)\,$.

 To keep the inclination-caused relative error at the level of $\,O(\inc^2)\,$, we still
 assume, in (\ref{32}), that $\,{\emph{l}}\,=\,2\,$, $\;m\,=\,2\,$, $\;p\,=\,0\,$. As for
 the the values of $\,q\,$, we keep only the ones giving us terms of order up to $\,e^4\,$,
 inclusively. Besides, we assume the phase lags to be small, so that
 $\,\sin\epsilon_{\textstyle{_{lmpq}}}\,=\,\epsilon_{\textstyle{_{lmpq}}}\,+\,O(\epsilon^3)
 \,=\,\epsilon_{\textstyle{_{lmpq}}}\,+\,O(Q^{-3})\,$. Under all these presumptions, the
 constant part of the tidal torque can be approximated with
 \begin{subequations}
 \label{627}
 \begin{eqnarray}
 \tau_{\textstyle{_{\textstyle{_{l=2}}}}}\;=\;
 \frac{3}{2}\;G\;M_{sat}^2\,R^5\,a^{-6}\,k_2\;\sum_{q=-2}^{2}\,
 G^{\textstyle{^{\,2}}}_{\textstyle{_{\textstyle{_{20\mbox{\it{q}}}}}}}\,
 \sin\epsilon_{\textstyle{_{\textstyle{_{220\mbox{\it{q}}}}}}}~+\,O(e^6/Q)\,
 +\,O(\inc^2/Q)\,~~~~~~~~~~~~~
 \label{627a}\\
 \nonumber\\
 =\;\frac{3}{2}\;G\;M_{sat}^2\,R^5\,a^{-6}\,k_2\;\sum_{q=-2}^{2}\,
 G^{\textstyle{^{\,2}}}_{\textstyle{_{\textstyle{_{20\mbox{\it{q}}}}}}}\,
 \epsilon_{\textstyle{_{\textstyle{_{220\mbox{\it{q}}}}}}}~+\,O(e^6/Q)\,
 +\,O(\inc^2/Q)\,+\,O(Q^{-3})\;\;\;,
 \label{627b}
 \end{eqnarray}
 \end{subequations}
 where, according to the tables (Kaula 1966),
 \ba
 \nonumber
 G^2_{\textstyle{_{\textstyle{_{20\;-2}}}}}\,=\,0\;\;\;,\;\;\;\;\;\;\;
 G^2_{\textstyle{_{\textstyle{_{20\;-1}}}}}\,=\,\frac{e^2}{4}\;-\;
 \frac{e^4}{16}\;+\;O(e^6)\;\;\;\,,\;\;\;\;
 G^2_{\textstyle{_{\textstyle{_{200}}}}}\,=\,1\,-\,5\,e^2\,+\;\frac{63}{8}\;e^4\;
 +\;O(e^6)\;\;\;\,,\;\;\;\;\;\\
 \label{628}\\
  \nonumber
 G^2_{\textstyle{_{\textstyle{_{20{{1}}}}}}}\,=\,\frac{49}{4}\;e^2\;-\;
 \frac{861}{16}\;e^4\;+\;O(e^6)\;\;\;\;,\;\;\;\;\;\;\;
 G^2_{\textstyle{_{\textstyle{_{20{{2}}}}}}}\,=\,\frac{289}{4}\,e^4\,+
 \,O(e^6)\;\;\;\;,~~~~~~~~~~~~~~~~~~~~~~~~~~~~~~~~~~~~\,
 \ea\\
 and, according to formula (\ref{24}),
 \ba
 \nonumber
 \epsilon_{\textstyle{_{\textstyle{_{220\;-2}}}}}=\,(-\,2\,\dot{\theta}\,)\;
 \Delta t_{\textstyle{_{\textstyle{_{220\;-2}}}}}\;\;\;,\;\;\;\;
 \epsilon_{\textstyle{_{\textstyle{_{220\;-1}}}}}=\,(-\,2\,\dot{\theta}\,+\,n)
 \;\Delta t_{\textstyle{_{\textstyle{_{220\;-1}}}}}\;\;\;,\;\;\;\;
 \epsilon_{\textstyle{_{\textstyle{_{2200}}}}}=\,(-\,2\,\dot{\theta}\,+\,2\,n)
 \;\Delta t_{\textstyle{_{\textstyle{_{2200}}}}}\;\;\;,\;\;\;\;\;\\
 \label{629}\\
 \nonumber
 \epsilon_{\textstyle{_{\textstyle{_{220{{1}}}}}}}\,=\,(-\,2\,\dot{\theta}\,+\,3\,n)
 \;\Delta t_{\textstyle{_{\textstyle{_{220{{1}}}}}}}\;\;\;\;,\;\;\;\;\;\;
 \epsilon_{\textstyle{_{\textstyle{_{220{{2}}}}}}}\,=\,(-\,2\,\dot{\theta}\,+\,4\,n)
 \;\Delta t_{\textstyle{_{\textstyle{_{220{{2}}}}}}}\;\;\;\;.~~~~~~~~~~~~~~~~~~~~~~
 ~~~~~~~~~~~~~~~~~~~~~~
 \ea
 Provided the quality factor scales as inverse frequency, all the time lags
 are the same constant $\,\Delta t\,$, so the above formulae all together entail,
 in the case of nonresonant prograde spin:
 \ba
 \nonumber
 \ddot{\theta}\;=\,\frac{\tau}{C}
 =\;
 {\cal K}\,
 \left[\;-\;\dot{\theta}\,\left(
 1\,+\,\frac{15}{2}\,e^2\,+\,\frac{105}{4}\,e^4\,+\,O(e^6)
 \right)\right.~~~~~~~~~~~~~~~~~~~~~~~~~~~~~~~~~~~~~~~~~~~~~~~\\
 \label{630}\\
 \nonumber
 +\;\left.\,n\,\left(1\,+\,\frac{27}{2}\,e^2\,+\,\frac{573}{8}\,e^4\,+\,
 O(e^6)\right) \,\right]~
 +\,O(\inc^2/Q)\,+\,O(Q^{-3})\;\;\;,~~~~~~~~~
 \ea
 which coincides with (\ref{647} - \ref{649}) to the order $\,e^4\,$,
 inclusively, provided we substitute $\,\chi_{\textstyle{_{2200}}}\,$ instead
 of $\,\chi\,$ in the expression (\ref{654}) for $\,{\cal{K}}\,$.

 \subsection{Rheologies different from
 $\,Q\,\sim\,1/\chibold\;\;$}

 A part and parcel of both afore-presented methods was the assertion of all the time lags
 $\,\Delta t_{\textstyle{_{lmpq}}}\,$ being equal. In reality, the time lags vary from one
 harmonic to another.

 Any particular functional form of the dependence $\,\Delta t(\chi)\,$ fixes the rheology:
 for example, the frequency-independence of $\,\Delta t\,$ constrains the value of the
 exponential $\,\alpha\,$ to $\,-1\,$ (while the parameter $\,{\cal{E}}\,$ becomes simply
 $\,\Delta t\,$). However, for an arbitrary $\,\alpha\,\neq\,-\,1\,$ the lags will read
 (Efroimsky \& Lainey 2007):
 \ba
 \Delta t_{\textstyle{_{\textstyle{_{lmpq}}}}}\;=\;{\cal{E}}^{\textstyle{^{-\,\alpha}}}\;
 \chi_{\textstyle{_{\textstyle{_{lmpq}}}}}^{\textstyle{^{-\,(\alpha+1)}}}
 \label{631}
 \ea
 While the MacDonald approach cannot be generalised to $\,\alpha\,\neq\,-\,1\,$,
 the Darwin-Kaula-Goldreich method can be well combined with (\ref{631}). To
 this end, we shall insert (\ref{628} - \ref{629}) and (\ref{631}) into
 (\ref{627a}), and shall also employ the evident formulae
 \ba
 \cos\epsilon_{\textstyle{_{\textstyle{_{{{{\it{l}}mpq}}}}}}}\,=\,\frac{~|\,\cot
 \epsilon_{\textstyle{_{\textstyle{_{{\it{l}}mpq}}}}}\,|\;}{\sqrt{{\textstyle
 1~+~\cot^2
 \epsilon_{\textstyle{_{\textstyle{_{{\it{l}}mpq}}}}}}}}=\;\frac{\,
 Q_{\textstyle{_{\textstyle{_{{\it{l}}mpq}}}}}\;}{\sqrt{{\textstyle 1~+~
 Q^{\textstyle{^{2}}}_{\textstyle{_{\textstyle{_{{\it{l}}mpq}}}}}}}}
 ~=~\frac{\;{\cal{E}}^{\textstyle{^{\alpha}}}\;
 \chi^{\textstyle{^{\alpha}}}_{\textstyle{_{\textstyle{_{{\it{l}}m
 pq}}}}}\;}{\sqrt{{\textstyle 1~+~
 {\cal{E}}^{\textstyle{^{2\alpha}}}\;
 \chi^{\textstyle{^{2\alpha}}}_{\textstyle{_{\textstyle{_{{\it{l}}mpq}}}}}}}}
 ~~~,~~~~~~~~~~~~~~~~~~~~
 \label{cos}
 \ea
 \ba
 \sin\epsilon_{\textstyle{_{\textstyle{_{{{{\it{l}}mpq}}}}}}}=\,
 \sin|\epsilon_{\textstyle{_{\textstyle{_{{\it{l}}mpq}}}}}|\;\,\mbox{sgn}\,
 \omega_{\textstyle{_{\textstyle{_{{\it{l}}mpq}}}}}=\,\frac{\mbox{sgn}\,
 \omega_{\textstyle{_{\textstyle{_{{\it{l}}mpq}}}}}\;}{\sqrt{{\textstyle 1~+~\cot^2
 \epsilon_{\textstyle{_{\textstyle{_{{\it{l}}mpq}}}}}}}}=\;\frac{\mbox{sgn}\,
 \omega_{\textstyle{_{\textstyle{_{{\it{l}}mpq}}}}}\;}{\sqrt{{\textstyle 1~+~
 Q^{\textstyle{^{2}}}_{\textstyle{_{\textstyle{_{{\it{l}}mpq}}}}}}}}
 ~=~\frac{\mbox{sgn}\,
 \omega_{\textstyle{_{\textstyle{_{{\it{l}}mpq}}}}}\;}{\sqrt{{\textstyle 1~+~
 {\cal{E}}^{\textstyle{^{2\alpha}}}\;
 \chi^{\textstyle{^{2\alpha}}}_{\textstyle{_{\textstyle{_{{\it{l}}mpq}}}}}}}}
 ~~~,~~~
 \label{sin}
 \ea
 with $\omega_{\textstyle{_{\textstyle{_{{\it{l}}mpq}}}}}\,$ given by (\ref{25}), and
 $\,|\epsilon_{\textstyle{_{\textstyle{_{{\it{l}}mpq}}}}}|\,$ assumed (for reasons explained in the Appendix)
 not to approach too close to
 $\,\pi/2\;$. This will give us the following expression for (the constant part of) the deceleration rate of a non-resonant prograde spin:
 \ba
 \nonumber
 \ddot{\theta}\,=\,\frac{\tau}{C}
 =\;-\;\frac{3}{2}\;\frac{G\;M^2_{sat}}{a^3}\,\frac{R^5}{a^3}\;\frac{k_{\textstyle{_2}}\;
 }{\xi\;M_{planet}\;R^2}
 \;\left[\;\frac{e^2}{4}\;\,\mbox{sgn}(2\,\dot{\theta}\,-\,n)\;\;
 \frac{{\cal{E}}^{\textstyle{^{\alpha}}}\,\;
 |2\,\dot{\theta}\,-\,n|^{\textstyle{^{\alpha}}}}{\;1\;+\;{\cal{E}}^{\textstyle{^{2\alpha}}}\,\;
 |2\,\dot{\theta}\,-\,n|^{\textstyle{^{2\alpha}}}\,}\;\; \right.~~~~~~~~~~~~~~~~~~~~~~~~~~~~~
  \label{expression}
  \ea
  \ba
  \nonumber
 \left.
 +\;\left(1-5e^2+\frac{63}{8}e^4\right)\;\mbox{sgn}(2\,\dot{\theta}\,-\,2\,n)\;\;
 \frac{{\cal{E}}^{\textstyle{^{\alpha}}}\,\;
 |2\,\dot{\theta}\,-\,2\,n|^{\textstyle{^{\alpha}}}}{\,1\;+\;{\cal{E}}^{\textstyle{^{2\alpha}}}\,\;
 |2\,\dot{\theta}\,-\,2\,n|^{\textstyle{^{2\alpha}}}\,}\;\right.~~~~~~\,
 \ea
 \ba
 \nonumber
 +\,\left.
 \left(\frac{49}{4}e^2-\frac{861}{16}e^4\right)\;\mbox{sgn}(2\,\dot{\theta}\,-\,3\,n)\;\;
 \frac{{\cal{E}}^{\textstyle{^{\alpha}}}\,\;
 |2\,\dot{\theta}\,-\,3\,n|^{\textstyle{^{\alpha}}}}{\,1\;+\;{\cal{E}}^{\textstyle{^{2\alpha}}}\,\;
 |2\,\dot{\theta}\,-\,3\,n|^{\textstyle{^{2\alpha}}}\,}
 \right.~~~~~~
 \ea
 \ba
 \left.~~~~~~~~~~~~~~~+\;
 \frac{289}{4}\;e^4\;\mbox{sgn}(2\,\dot{\theta}\,-\,4\,n)\;
 \frac{{\cal{E}}^{\textstyle{^{\alpha}}}\,\;
 |2\,\dot{\theta}\,-\,4\,n|^{\textstyle{^{\alpha}}}}{\,1\;+\;{\cal{E}}^{\textstyle{^{2\alpha}}}\,\;
 |2\,\dot{\theta}\,-\,4\,n|^{\textstyle{^{2\alpha}}}\,}
 \;\right]\;+\;O(\inc^2/Q)\;+\;O(e^6/Q)~~~.~~~~~~~
 \label{formula}
 \ea
 Be mindful, that a naive substitution of the formula (\ref{sin}) for $\,\sin
 \epsilon{\textstyle{_{\textstyle{_{{\it{l}}mpq}}}}}$ into (\ref{627a}) would
 result in an expression for the torque, attaining its maxima on approach to
 resonances (for a positive $\alpha$), an evidently unphysical behaviour.
 As explained in section 9, there exists a profound physical reason,
 for which the actual multiplier in (\ref{627a}) must be not $\,\sin
 \epsilon{\textstyle{_{\textstyle{_{{\it{l}}mpq}}}}}\;$ but: $\;\sin
 \epsilon{\textstyle{_{\textstyle{_{{\it{l}}mpq}}}}}\,
 \cos\epsilon{\textstyle{_{\textstyle{_{{\it{l}}mpq}}}}}\,$.
 Mathematically, the presence of the cosine is irrelevant unless $\chi_{\textstyle{_{\textstyle{_{{\it{l}}mpq}}}}}$
 and $Q_{\textstyle{_{\textstyle{_{{\it{l}}mpq}}}}}$ approach zero. If
 however $\,\chi_{\textstyle{_{\textstyle{_{{\it{l}}mpq}}}}}\,$ becomes very
 small (i.e., if we approach a resonance), it is this long-omitted (though known
 yet to Darwin 1879) cosine multiplier that saves us from the unphysical maxima
 -- see section 9 above.

 Under the extra assumptions\footnote{~The ~smallness ~of
 $\;\,|\,\epsilon_{\textstyle{_{\textstyle{_{{\it{l}}mpq}}}}}|\;\,$ enables
 ~one ~to ~employ ~(\ref{627b}) ~instead ~of ~(\ref{627a}). ~~Then ~the
 ~multipliers\\
 $
 {\left.\;~\right.}^{\left.\;~\right.}\\
 \frac{\mbox{sgn}\;\omega_{\textstyle{_{\textstyle{_{{\it{l}}mpq}}}}}\,
 \;\,\textstyle{{\cal E}^{\textstyle{^\alpha}}\,
 \chi_{\textstyle{_{\textstyle{_{{\it{l}}mpq}}}}}^{\textstyle{^\alpha}}}}{
 \textstyle{1+{\cal E}^{\textstyle{^{2\alpha}}}\,
 \chi_{\textstyle{_{\textstyle{_{{\it{l}}mpq}}}}}^{\textstyle{^{2\alpha}}}}}
 \,\;$
  in (\ref{formula})
 become
 $\;\,\epsilon_{\textstyle{_{\textstyle{_{{\it{l}}mpq}}}}}
 =
 \,\omega_{\textstyle{_{\textstyle{_{{\it{l}}mpq}}}}}\Delta t_{\textstyle{_{\textstyle{_{2200}}}}}
 \frac{\textstyle{\Delta t_{\textstyle{_{\textstyle{_{{\it{l}}mpq}}}}}}}{\textstyle{\Delta t_{\textstyle{_{\textstyle{_{2200}}}}}}}
 =
\chi_{\textstyle{_{\textstyle{_{{\it{l}}mpq}}}}}\,
 \Delta t_{\textstyle{_{\textstyle{_{2200}}}}} \,
\mbox{sgn}\;\omega_{\textstyle{_{\textstyle{_{{\it{l}}mpq}}}}}
\left(\frac{\textstyle{\chi_{\textstyle{_{\textstyle{_{2200}}}}}}}{
\textstyle{\chi_{\textstyle{_{\textstyle{_{{\it{l}}mpq}}}}}}}\right)^{\textstyle{^{\alpha+1}}}\\
{\left.\;~\right.}^{\left.\;~\right.}\\
 = \chi_{\textstyle{_{\textstyle{_{2200}}}}}\,
 \Delta t_{\textstyle{_{\textstyle{_{2200}}}}} \,
\mbox{sgn}\;\omega_{\textstyle{_{\textstyle{_{{\it{l}}mpq}}}}}
\left(\frac{\textstyle{\chi_{\textstyle{_{\textstyle{_{2200}}}}}}}{
\textstyle{\chi_{\textstyle{_{\textstyle{_{{\it{l}}mpq}}}}}}}\right)^{\textstyle{^{\alpha}}}
 =\chi_{\textstyle{_{\textstyle{_{2200}}}}}\,
 \Delta t_{\textstyle{_{\textstyle{_{2200}}}}} \,
 \mbox{sgn}\;\omega_{\textstyle{_{\textstyle{_{{\it{l}}mpq}}}}}
 \left(1\; +\;
 \frac{\textstyle{\chi_{\textstyle{_{\textstyle{_{{\it{l}}mpq}}}}}}\;-\;
 \textstyle{\chi_{\textstyle{_{\textstyle{_{2200}}}}}}}{
\textstyle{\chi_{\textstyle{_{\textstyle{_{2200}}}}}}}
\right)^{\textstyle{^{-\,\alpha}}}.~~~$ Specifically,
$ {\left.\;~\right.}^{\left.\;~\right.}\\
 {\left.\;~\right.}^{\left.\;~\right.}\\
 \epsilon_{\textstyle{_{\textstyle{_{220{\textstyle{q}}}}}}}=\mbox{sgn}\;\omega_{
 \textstyle{_{\textstyle{_{220{\textstyle{q}}}}}}}\;
 \Delta t_{\textstyle{_{\textstyle{_{2200}}}}}~
 \chi_{\textstyle{_{\textstyle{_{2200}}}}}
 \left(1\,+\,\frac{\textstyle{
 \chi_{\textstyle{_{\textstyle{_{{{{2}}20{\textstyle{q}}}}}}}}
 -\chi_{\textstyle{_{\textstyle{_{2200}}}}}}}{\textstyle{
 \chi_{\textstyle{_{\textstyle{_{2200}}}}}}}\right)^{\textstyle{^{-\,\alpha}}}
 \approx\mbox{sgn}\;
 \omega_{\textstyle{_{\textstyle{_{{{2}}20{\textstyle{q}}}}}}}
 \;\Delta t_{\textstyle{_{\textstyle{_{2200}}}}}\;
 \chi_{\textstyle{_{\textstyle{_{2200}}}}}\left(1\,-\,\alpha\;
 \frac{\textstyle{\chi_{\textstyle{_{\textstyle{_{{{{2}}20{\textstyle{q}}}}}}}}
 -\chi_{\textstyle{_{\textstyle{_{2200}}}}}}}{\textstyle{
 \chi_{\textstyle{_{\textstyle{_{2200}}}}}}}\right)\\
  {\left.\;~\right.}^{\left.\;~\right.}\\
  {\left.\;~\right.}^{\left.\;~\right.}\\
=\mbox{sgn}\;\omega_{\textstyle{_{\textstyle{_{{{2}}20{\textstyle{q}}}}}}}
\; \Delta t_{\textstyle{_{\textstyle{_{2200}}}}}
\;\left[\,(1\,+\,\alpha)\,\chi_{\textstyle{_{\textstyle{_{2200}}}}}\,
-\,\alpha\;\textstyle{
\chi_{\textstyle{_{\textstyle{_{{{{2}}20{\textstyle{q}}}}}}}}}
 \right]
\,\;$,
the latter approximation being legitimate only under the\\
 ${\left.\;~\right.}^{\left.\;~\right.}$\\  condition of
 $\, \chi_{\textstyle{_{\textstyle{_{220{\textstyle{q}}}}}}}\,-\,
 \chi_{\textstyle{_{\textstyle{_{2200}}}}}\,\ll\,
 \chi_{\textstyle{_{\textstyle{_{2200}}}}}\,$, which turns out to be equivalent
 to $\,n\,\ll\,\dot{\theta}\,$. For example,\\
 ${\left.\;~\right.}^{\left.\;~\right.}$\\
 $\;
 \frac{\chi_{\textstyle{_{\textstyle{_{220{\textstyle{q}}}}}}}\,-\,
 \chi_{\textstyle{_{\textstyle{_{2200}}}}}}{
 \chi_{\textstyle{_{\textstyle{_{2200}}}}}}\,=\,\frac{\textstyle(\,-\,2\,
 \stackrel{\centerdot}{\theta}\,+\,4\,n)\,-\,(\,-\,2\,\stackrel{\centerdot}{
 \theta}\,+\,2\,n)}{\textstyle{\,-\,2\,\stackrel{\centerdot}{\theta}\,+\,2\,n
 }}\,=\,\frac{\textstyle 2\,n}{\textstyle\,-\,2\,\stackrel{\centerdot}{\theta}
 \,+\,2\,n}\;\,$.
 So approximating
 $\,\epsilon_{\textstyle{_{\textstyle{_{220{\textstyle{q}}}}}}}\,$, for\\
 ${\left.\;~\right.}^{\left.\;~\right.}$\\
 $\,q\,=\,-\,2\,,\,-\,1\,,\,0\,,\,1\,,\,2\,$, ~we arrive at formula (\ref{632}).

 There exists one more reason to keep $\,n\,$ much smaller than $\,\dot{\theta}\,$ in
(86 - 89). We derived (86 - 89) by inserting the customary relation (36)
into (84 - 85). As explained in the Appendix below, (36) becomes invalid
near spin-orbit commensurabilities. Indeed, at each commensurability a
certain tidal harmonic becomes nil -- see formula (26). According to (60),
the appropriate Q, too, becomes nil. In this situation, one has to rely
not on (36) but on a more general formula (105). The latter formula
however entails vanishing of the appropriate  component of the tidal
torque on crossing the commensurability -- see (113 - 114). This is why in
(87 - 89) and even earlier, in (86), we should stay away from the
commensurabilities $\,\dot{\theta}=n/2\,$, $\,\dot{\theta}=n\,$,
$\,\dot{\theta}=3n/2\,$, or $\,\dot{\theta}=2n\,$. So we better keep
$\,n\ll \dot{\theta}\,$.}
 of $\,|\epsilon{\textstyle{_{\textstyle{_{{\it{l}}mpq}}}}}|\,\ll\,1\,$
 and $\,n\,\ll\,\dot{\theta}\,$, formula (\ref{formula}) simplifies  to
 \ba
 \nonumber
 \ddot{\theta}\,=\,\frac{\tau}{C}
 =
 {\cal{K}}
 \left[\;\,-\;\,\dot{\theta}\,\left(
 1\,+\,\frac{15}{2}\,e^2\,+\,\frac{105}{4}\,e^4\,+\,O(e^6)
 \right)\right.
 ~~~~~~~~~~~~~~~~~~~~~~~~~~~~~~~~~~~~~~~~~~~~~~~~~~~~~~~~~~~~~~
 \\
 \nonumber\\
 \nonumber\\
 \label{632}
 +\;\left.n\left(1+\left(\frac{15}{2}-6\alpha\right)e^2+
 \left(\frac{105}{4}-\frac{363}{8}\alpha\right)e^4+O(e^6)\right)\right]
 +O(\inc^2/Q)+O(Q^{-3})+O(n/\dot{\theta})\;\;\;~~~~~~
 \\
 \nonumber\\
 \nonumber\\
 \approx\;
 {\cal K}\;
 \left[\,-\,\dot{\theta}\,\left(
 1\,+\,\frac{15}{2}\,e^2\right)
  +\,n\,\left(1\,+\,\left(\frac{15}{2}\,-\,6\,\alpha\right)\,e^2\,\right)\,\right]\;\;\;,
  ~~~~~~~~~~~~~~~~~~~~~~~~~~~~~~~~~~~~~~~~~~
 \label{633}
 \ea
 where the overall factor ${\cal K}$ is given by
 \ba
 {\cal K}\,=\,\frac{3\,n^2\,M_{sat}^{\textstyle{^{\,2}}}\;\,k_{2}\;{\Delta t_{\textstyle{_{2200}}}}}{\xi\;
 M_{planet}\;(M_{planet}\,+\,M_{sat})}\;\frac{R^{\textstyle{^3}}}{a^3}
 \,=\,\frac{3\,n\,M_{sat}^{\textstyle{^{\,2}}}\;\,k_{2}}{\xi\;
 Q_{\textstyle{_{\textstyle{_{2200}}}}}\;M_{planet}\;(M_{planet}\,+\,M_{sat})}
 \;\frac{R^{\textstyle{^3}}}{a^3}\;\,
 \frac{n~~}{\chi_{\textstyle{_{\textstyle{_{2200}}}}}}\;\;\;,~~~~~~~~~~
 \label{634}
 \ea
  an expression identical to (\ref{654}), except that it contains
 $\,\Delta t_{\textstyle{_{2200}}}\,$, $\,Q_{\textstyle{_{2200}}}\,$, and
 $\,\chi_{\textstyle{_{2200}}}\,$ instead of $\,\Delta t\,$, $\,Q\,$, and $\,\chi\,$,
 correspondingly.

 Were $\;\alpha\;$ equal to $\;\,-\,1\;$, ~sum (\ref{632}) would coincide with ~(\ref{630}),
 provided $\;\dot{\theta}\,>\,2\,n\;$ (but not  otherwise!).
 For realistic mantles and crusts, though, the values of $\,\alpha\,$ will, as pointed above,
 reside within the interval $\,0.2 - 0.4\;$ (closer to $0.2$ for partial melts).

\pagebreak

 \section{Conclusions}

 In the article thus far we have provided a detailed review of a narrow
 range of topics. Our goal was to punctiliously spell out the assumptions
 that often remain implicit, and to bring to light those steps in
 calculations, which are often omitted as ``self-evident".

 This has helped us to demonstrate that MacDonald-style formula (\ref{45})
 for the tidal torque is valid only in the zeroth order of $\,en/Q\chi\,$,
 while its time-average is valid only in the first order. These restrictions
 mean that in the popular expressions for tidal despinning rate the terms
 with higher powers of $\,e\,$ become significant only for large eccentricities.
 Their significance is limited even further by the error $\,O(Q^{-3})\,$
 emerging when the sine of the phase lag gets approximated by the inverse
 quality factor -- see formula (\ref{46}) and the paragraph preceding it.

 We have demonstrated that in the case, when the inclinations are small and the
 phase lags of the tidal harmonics are proportional to the frequency, the
 Darwin-Kaula expansion is equivalent to a corrected version of the MacDonald
 formalism. The latter method rests on the assumption of existence of one total
 double bulge. The necessary correction to MacDonald's approach would be to assert
 (following Singer 1968) that the phase lag of this integral bulge is not
 constant, but is proportional to the instantaneous synodal frequency
 $\,2(\dot{\nu}-\dot{\theta})\,$, where $\nu$ and $\theta$ are the true anomaly
 and the sidereal angle. Any rheology different from this one will violate the
 equivalence of the Darwin-Kaula and MacDonald descriptions. It remains
 unexplored if their equivalence is violated also by setting the inclination high.

 We have demonstrated that no ``paradoxes" ensue from the frequency-dependence
 $\,Q\sim\chi^{\alpha}\;$, with $\;\;\alpha\,=\,0.3\,\pm\,0.1\;$, found for the
 mantle.

 We have investigated the limitations of the popular formula interconnecting
the quality factor $\,Q\,$ and the phase lag $\,\epsilon\,$. It turns out that
 for low quality factors (less than 10),
the customary formula $\,Q\, = \,\cot |\epsilon|\,$ should be substituted with a far more complicated
relation.

 Finally, we examined two derivations of the popular expressions (\ref{Laskar} -
 \ref{649}), and have pointed out that these expressions have limitations
 related to the frequency-dependence of the quality factor. First, dependent upon
 the values of $e$ and $Q$, the high-order terms in these expressions may become
 significant only for large eccentricities. Second, the expansion of the
 deceleration rate in even powers of $\,e\,$ will be different if $\,\Delta t\,$
 is frequency-dependent (which is the case for solid materials). These two
 circumstances do not necessarily disprove any major result achieved in the
 bodily-tide theory. However, some coefficients may now have to be reconsidered.

 For the realistic rheology of terrestrial bodies, the despinning rate, in the
 absence of tidal locking, is given by our formulae (\ref{expression} -
 \ref{634}).

 ~\\

 {\underline{\textbf{\Large{Acknowledgments}}}}\\
 ~\\
 It is a pleasure for us to acknowledge the contribution to this work from Alessandra
 Celletti, whose incisive questions ignited a discussion and eventually compelled us
 to take pen to paper. Our profoundest gratitude goes also to Sylvio Ferraz Mello, who
 kindly offered a large number of valuable comments and important corrections. ME would
 also like deeply to thank Bruce Bills, Tony Dobrovolskis, Peter Goldreich, Shun-ichiro
 Karato, Valery Lainey, William Newman, Stan Peale, S. Fred Singer, Victor Slabinski, and
 Gabriel Tobie for numerous stimulating conversations on the theory of tides. Part of the research described in this paper was carried out at the Jet Propulsion Laboratory of the
 California Institute of Technology, under a contract with the National Aeronautics and
 Space Administration. ME is most grateful to John Bangert for his support of the project on all
 of its stages.

 \pagebreak

  \noindent
  {\underline{\textbf{\Large{Appendix.}}}}\\
  ~\\
  {\textbf{\large{The$\,$lag~and~the~quality~factor:~is~the~formula~$\boldmath{Q=\cot |\epsilonbold |}$~universal?}}}\\

  The interrelation between the quality factor $\,Q\,$ and the phase lag $\,\epsilon\,$ is
 long-known to be
 \ba
 Q\,=\,\cot |\epsilon |\;\;\;,
 \label{formulla}
 \ea
 and its derivation can be found in many books. In Appendix A2 of Efroimsky \& Lainey(2007),
 that derivation is reproduced, with several details that are normally omitted in the literature. Among
 other things, we pointed out that the interrelation has exactly the form (\ref{formulla})
 only in the limit of small lags. For large phase lags, the form of this relation will
 change considerably.

 Since in section 9 of the current paper we address the case of large lags, it would be worth reconsidering
 the derivation presented in Efroimsky \& Lainey (2007), and correcting a subtle omission made there. Before
 writing formulae, let us recall that, at each frequency $\,\chi\,$ in the spectrum of the deformation, the
 quality factor (divided by $\,2\,\pi\,$) is defined as the peak energy stored in the system divided by the
 energy damped over a cycle of flexure:
 \ba
 {Q}(\chi)\;\equiv\;-\;\frac{2\;\pi\;E_{peak}(\chi)}{\Delta E_{cycle}(\chi)}\;\;\;,
 \label{}
 \ea
 where $\,\Delta E_{cycle}(\chi)\,<\,0\,$ as we are talking about energy
 losses.\footnote{~We are considering flexure in the linear approximation. Thus at each
  frequency $\,\chi\,$ the appropriate energy loss over a cycle, $\,\Delta E_{cycle}(\chi)\,$, depends solely on the maximal energy stored at that same frequency, $\,E_{peak}(\chi)\,$.}

 An attempt to
 consider large lags (all the way up to $\,|\epsilon |\,=\,\pi/2\,$) sets the values of $\,Q/2\pi\,$
 below unity. As the dissipated energy cannot exceed the energy stored in a free oscillator, the
 question becomes whether the values of $\,Q/2\pi\,$ can be that small. To understand
 that they can, recall that in this situation we are considering an oscillator, which is not
 free but is driven (and is overdamped). The quality factor being much less than unity simply
 implies that the eigenfrequencies get damped away during less than one oscillation. Nonetheless,
 motion goes on due to the driving force.

 Now let us switch to the specific context of tides.  To begin with, let us recall that the dissipation
 rate in a tidally distorted primary is well approximated by the work that the
 secondary carries out to deform the primary:
 \ba
 \dot{E}\;=\;-\;\int \,\rho\;\Vbold\;\cdot\;\nabla W\;d^3x
 \label{A9}
 \ea
 $\rho\,,\;\Vbold\,$, and $\,W\,$ denoting the density, velocity, and tidal
 potential inside the primary. The expression on the right-hand side can be
 transformed by means of the formula
 \ba
 \rho\,\Vbold\cdot\nabla W\,=\,\nabla\cdot(\rho\,\Vbold\,W)\,-\,W\,\Vbold\cdot
 \nabla\rho\,-\,W\,\nabla\cdot(\rho\,\Vbold)\,=\,\nabla\cdot(\rho\,
 \Vbold\,W)\,-\,W\,\Vbold\cdot\nabla\rho\,+\,W\,
 \frac{\partial\rho}{\partial t}\;\;,\;\;\;\;
 \label{}
 \ea
 where the $W\Vbold \cdot \nabla \rho$ and $\partial\rho/\partial t$ terms may
 be omitted under the assumption that the primary is homogeneous and incompressible.
 In this approximation, the attenuation rate becomes simply
 \ba
 \dot{E}\;=\;-\;\int\,\nabla\,\cdot\,(\rho\;\Vbold\;W)\,d^3x
 \;=\;-\;\int\,\rho\;W\;\Vbold\,\cdot\,{\vec{\bf{n}}}\;\,dA\;\;\;,
 \label{}
 \ea
 ${\vec{\bf{n}}}\,$ being the outward normal to the surface of the primary,
 and $\,dA\,$ being an element of the surface area. It
 is now clear that, under the said assertions, it is sufficient to take into
 account only the radial elevation rate, not the horizontal distortion. This
 way, formula (\ref{A9}), in application to a unit mass, will get simplified to
 \ba
 \dot{E}\;=\;\left(-\,\frac{\partial W}{\partial r}\right)\;\Vbold\cdot{\vec{\bf{n}}}
 \;=\;\left(-\,\frac{\partial W}{\partial r}\right)\frac{d\zeta}{dt}\;\;\;,
 \label{}
 \ea
 $\zeta\,$ standing for the vertical displacement (which is, of course, delayed
 in time, compared to $\,W\,$). The amount of energy dissipated over a time
 interval $\,(t_o\,,\;t)\,$ is then
 \ba
 \Delta{E}\;=\;\int^{t}_{t_o}\;\left(-\,\frac{\partial W}{\partial
 r}\right)\;d\zeta\;\;\;.
 \label{dissipation}
 \ea

 We shall consider the simple
 case of an equatorial moon on a circular orbit. At each point of the planet, the
 variable part of the tidal potential produced by this moon will read
  \ba
  W\;=\;W_o\;\cos \chi t\;\;\;,
  \label{A3}
  \label{469}
  \ea
 the tidal frequency being given by
 \ba
 \chi\,=\,2~|n\;-\;\omega_p|~~~.~~~
 \label{A3}
 \label{470}
 \ea
 Let $\,\mbox{g}\,$ denote the surface free-fall acceleration. An element of the planet's surface lying beneath the satellite's trajectory will then experience a vertical elevation  of
 \ba
 \zeta\;=\;h_2\;\frac{W_o}{\mbox{g}}\;\cos (\chi t\;-\;|\epsilon |)\;\;\;,
 \label{A4}
 \label{471}
 \ea
 $\,h_2\,$ being the corresponding Love number\footnote{~For a homogeneous incompressible
 body, $\,k_2 = (3/5) h_2\,$, for which reason (\ref{471}) and the subsequent equations with $\,h_2\,$ can equally be written as proportional to $\,k_2\,$. The formulation employing
 $\,k_2\,$ is more fundamental, as it can, in principle, be generalised to a compressible
 body of a radially-changing density. Indeed, whatever the properties of the primary
 are, the dissipation rate in it is equal to the rate of change of the primary's spin energy plus the rate of change of the orbital energy. Both the latter and the former
 are proportional to $\,k_2\,$.}, and $\,|\epsilon |\,$ being the
 {\emph{positive}}\footnote{~Were we not considering the simple case of a circular orbit, then, rigorously
 speaking, the expression for $\,W\,$ would read not as $\,W_o\,\cos \chi t\,$ but
 as $\,W_o\,\cos \omega_{\textstyle{_{tidal}}} t\,$, the tidal frequency $\,\omega_{\textstyle{_{tidal}}}\,$ taking both positive and negative
 values, and the physical frequency of flexure being $\,\chi\,\equiv\,|\omega_{\textstyle{_{tidal}}}|\,$.
 Accordingly, the expression for $\,\zeta\,$ would contain not $\,\cos (\chi t\,-\,|\epsilon |)\,$ but
 $\,\cos (\omega_{\textstyle{_{tidal}}} t\,-\,\epsilon)\,$. As we saw in equation (\ref{24}), the sign of $\,\epsilon\,$
 is always the same as that of $\,\omega_{\textstyle{_{tidal}}}\,$. For this reason, one may simply deal with
 the physical frequency $\,\chi\,\equiv\,|\omega_{\textstyle{_{tidal}}}|\,$ and with the absolute value of the phase lag, $\;| \epsilon |\;$.}
 phase lag, which for the principal tidal frequency is simply the double
 geometric angle $\,\delta\,$ subtended at the primary's centre between the
 directions to the secondary and to the main bulge:
 \ba
 |\epsilon|\;=\;2\;\delta\;\;\;.
 \label{dot}
 \ea
 Accordingly, the vertical velocity of this element of the planet's surface will
 amount to
  \ba
 u\;=\;\dot{\zeta}\;=\;-\;h_2\;\chi\;\frac{W_o}{\mbox{g}}\;\sin (\chi t
 \;-\;|\epsilon|)\;=\;-\;h_2\;\chi\;\frac{W_o}{\mbox{g}}\;\left(\sin \chi
 t\;\cos |\epsilon|\;-\;\cos \chi t\; \sin |\epsilon|\right)\;\;.\;\;
 \label{A5}
 \label{472}
 \ea
 The expression for the velocity has such a simple form because in this case the
 instantaneous frequency $\chi$ is constant. The satellite generates two bulges
 (on the facing and opposite sides of the planet) so each point of the surface
 is uplifted twice through a cycle. This entails the factor of two in the
 expression (\ref{470}) for the frequency. The phase in (\ref{dot}), too, is
 doubled, though the necessity of this is less evident, -- see footnote 4 in
 Appendix to Efroimsky \& Lainey (2007).

 The energy dissipated over a time cycle $\,T\,=\,2\pi/\chi\,$, per
 unit mass, will, in neglect of horizontal displacements, be
 \ba
 \nonumber
 \Delta E_{_{cycle}} &=& \int^{T}_{0}u\left(-\,\frac{\partial W}{
 \partial r}\right)dt=
 \,-\left(-\,h_2\;\chi \frac{W_o}{\mbox{g}}\right)\,\frac{\partial W_o}{
 \partial r}\int^{t=T}_{t=0}\cos \chi t\,\left(\sin \chi t\,
 \cos |\epsilon|\,-\,\cos \chi t\, \sin |\epsilon|\right)dt\\
 \nonumber\\
 \nonumber\\
 &=&\,-\;h_2\;\chi\;\frac{W_o}{\mbox{g}}\;\frac{\partial W_o}{\partial r}\;\sin|\epsilon|\,
 \;\frac{1}{\chi}\;\int^{\chi t\,=\,2\pi}_{\chi t\,=\,0}\;\cos^2 \chi t\;\;d(\chi
 t)\;=\;-\;h_2\;\frac{W_o}{\mbox{g}}\;\frac{\partial W_o}{\partial r}\;\pi\;\sin|\epsilon|
 \;\;,\;\;\;~~~~~~~~~~~~~~~~~~~~
 \label{A6}
 \label{}
 \ea
 while the peak energy stored in the system during the cycle will read:
 \ba
 \nonumber
 E_{_{peak}}&=&\int^{T/4}_{|\epsilon|/\chi} u \left(-\,\frac{\partial W}{
 \partial r}\right)dt =
 \,-\left(-\,h_2\;\chi\,\frac{W_o}{\mbox{g}}\right)\frac{\partial W_o
 }{\partial r}\int^{t=T/4}_{t=|\epsilon|/\chi}\cos \chi t\,\left(\sin
 \chi t\,\cos |\epsilon|\,-\,\cos \chi t\,\sin |\epsilon|\right)dt\\
 \nonumber\\
 \nonumber\\
 &=&\;\chi\;h_2\;\frac{W_o}{\mbox{g}}\;\frac{\partial W_o}{\partial r}\;\left[\;
 \frac{\cos |\epsilon|}{\chi}\;\int^{\chi t\,=\,\pi/2}_{\chi t\,=\,|\epsilon|}
 \;\cos \chi t\;\sin \chi t\;\;d(\chi t)\;-\;\frac{\sin |\epsilon|
 }{\chi}\;\int^{\chi t\,=\,\pi/2}_{\chi t\,=\,|\epsilon|}\;\cos^2 \chi t
 \;\;d(\chi t)\;\right]\;\;.~~~~~~~~\,
 \ea
 In the appropriate expression in Appendix A1 to Efroimsky \& Lainey (2007),
 the lower limit of integration was erroneously set to be zero. To understand that in
  reality integration over $\chi t$ should begin from $|\epsilon|$, one should superimpose the
  plots of the two functions involved, $\cos \chi t$ and $\sin(\chi t-|\epsilon|)$. The maximal energy
  gets stored in the system after integration through the entire interval over which both
  functions have the same sign. Hence $\chi t=|\epsilon|$ as the lower limit.

 Evaluation of the integrals entails:
 \ba
 E_{peak}\;=\;h_2\;\frac{W_o}{\mbox{g}}\;\frac{\partial W_o}{\partial r}\;\left[\;\frac{1}{2}
 \;\cos |\epsilon|\;-\;\frac{1}{2}\;\left(\;\frac{\pi}{2}\;-\;|\epsilon|\;\right)\;\sin |\epsilon|\;\right]~~~~~~~~~~~~~~~~~~~
 ~~~~~~~~~~~~~~~~~~~~~~~~~~~~~~~~~~~~~~~~~~~~~~~~~~~~~~~
 \label{A7}
 \label{}
 \ea
 whence
 \ba
 Q^{-1}\;=\;\frac{-\;\Delta E_{_{cycle}}}{2\,\pi\,E_{_{peak}}}\;=\;\frac{1}{2\,\pi}
 \;\,\frac{\pi\;\sin |\epsilon|}{~\frac{\textstyle 1}{\textstyle 2}\;\cos |\epsilon|\;-\;
 \frac{\textstyle 1}{\textstyle 2}\;\left(\;\frac{\textstyle \pi}{\textstyle 2}\;-\;
 |\epsilon|\;\right)\;\sin |\epsilon|}\;=\;\frac{\tan |\epsilon|}{1\;-\;\left(\;\frac{\textstyle \pi}{\textstyle 2}\;-\;
 |\epsilon|\;\right)  \;\tan|\epsilon|}\;\;\;.~~~~~
 \label{A8}
 \label{}
 \ea
 As can be seen from (\ref{A8}), both the product $\,\sin\epsilon\,\cos\epsilon\,$ and
 the appropriate component of the torque attain their maxima when
 $\,Q\,=\,1\,-\,\pi/4\,$.

 Usually, $\,|\epsilon|\,$ is small, and we arrive at the customary expression
 \ba
 Q^{-1}\,=\,\tan|\epsilon|\;+\;O(\epsilon^2)\;\;\;.
 \label{customary}
 \ea
 In the opposite case, when $Q\rightarrow 0$ and $|\epsilon| \rightarrow \pi/2$,
 it is convenient to employ the small difference
 \ba
 \xi\;\equiv\;\frac{\pi}{2}\;-\;|\epsilon|\;\;\;,
 \label{}
 \ea
 in terms whereof the inverse quality factor will read:
 \ba
 Q^{-1}\,=\,\frac{\cot \xi}{1\;-\;\xi
 \;\cot\xi}\;=\;\frac{1}{\tan\xi\;-\;\xi}\;=\;\frac{1}{
 z\;-\;\arctan z}\;=\;\frac{1}{\frac{\textstyle 1}{\textstyle 3}\;z^3\;\left[
 \,1\;-\;\frac{\textstyle 3}{\textstyle 5}\;z^2\,+\;O(z^4)\,\right]}
 \;\;\;,\;\;\;
 \label{A69}
 \ea
 where $\;z\,\equiv\,\tan\xi\;$ and, accordingly,
 $\;
 \xi\;=\;\arctan z\;=\;z\,-\frac{\textstyle 1}{\textstyle 3}\,z^3\,+\,\frac{\textstyle 1}{\textstyle 5}\,z^5\,+\,O(z^7)\;\,.
 \;$
 Formula (\ref{A69}) may, of course, be rewritten as
 \ba
 z^3\;\left[\,1\;-\;\frac{3}{5}\;z^2\;+\;O(z^4)\;\right]\;=\;3\;Q
 \;\;\;
 \label{}
 \ea
 or, the same, as
 \ba
 z\;=\;(3\,Q)^{1/3}\;\left[\,1\;+\;\frac{1}{5}\;z^2\;+\;O(z^4)\,\right]\;\;\;.
 \label{}
 \ea
 While the zeroth approximation is simply $\;z\,=\,(3Q)^{1/3}\,+\,O(Q)\;$, the first
 iteration gives:
 \ba
 \tan\xi\;\equiv\;z\;=\;
 (3Q)^{1/3}\,\left[\,1\;+\;\frac{1}{5}\;(3Q)^{2/3}\;+\;O(Q^{4/3})\,\right]\;
 =\;q\;\left[\,1\;+\;\frac{1}{5}\;q^2\;+\;O(q^4)\,\right]\;\;\;,~~~
 \label{}
 \ea
 with $\,q\,=\,(3Q)^{1/3}\,$ playing the role of a small parameter.

 We now see that the customary relation (\ref{customary}) should be substituted, for
 large lags, i.e., for small\footnote{~The afore-employed expansion of $\,\arctan z\,$ is valid for
 $\,|z|\,<\,1\,$. This inequality, along with (\ref{A69}), entails: $\,Q\,=\,z\,-\,\arctan z\,<\,1\,-\,\pi/4\,$.}
 values of $\,Q\,$, with:
 \ba
 \tan|\epsilon|\;=\;(3Q)^{-1/3}\,\left[\,1\;-\;\frac{1}{5}\;(3Q)^{2/3}\;+\;O(Q^{4/3})\,\right]
 \label{}
 \ea
 The formula for the tidal torque contains a multiplier
 $\,\sin \epsilon\,\cos\epsilon\,$, whose absolute value can, for our purposes, be written down as
 \ba
 \sin|\epsilon|\,\cos|\epsilon|=\cos\xi\,\sin\xi=\frac{\tan\xi}{1+\tan^2\xi}
 \,=\,\frac{q\,\left[1+\frac{\textstyle 1}{\textstyle 5}\,q^2+O(q^4)\right]}{1+q^2\left[1
 +O(q^2)\right]}
  =(3Q)^{1/3}\left[1-\frac{4}{5}(3Q)^{2/3}+O(Q^{4/3})\right]\;,~~~
 \label{}
 \ea
 whence
 \ba
 \sin\epsilon\;\cos\epsilon\;=\;\pm\;(3Q)^{1/3}\left[1-\frac{4}{5}(3Q)^{2/3}+O(Q^{4/3})\right]\;,~~~
 \label{}
 \ea
 an expression vanishing for $\,Q\,\rightarrow\,0\;$. Be mindful that both $\,\epsilon_{\textstyle{_{2200}}}\,$
 and the appropriate component of the torque change their sign on the satellite crossing the synchronous
 orbit.

 \end{document}